\documentclass[a4paper,fleqn]{cas-dc}

\usepackage[authoryear]{natbib}
\usepackage{graphicx}
\usepackage{ import } % Figure folder
\usepackage{ multirow }
\usepackage{ caption, subcaption }
\usepackage{listings} % Code snippet
\usepackage{xspace} % Matlab symbol
\usepackage{xpatch} % Code inline
\usepackage{xcolor}
\usepackage{realboxes}
\definecolor{mygray}{rgb}{0.8,0.8,0.8}
\makeatletter
\xpretocmd\lstinline{\Colorbox{mygray}\bgroup\appto\lst@DeInit{\egroup}}{}{}
\makeatother

\SetSymbolFont{letters}{bold}{OML}{cmm}{b}{it}
\SetSymbolFont{operators}{bold}{OT1}{cmr}{bx}{n}

\newcommand{\lpare}[1]{\left(#1\right)} % Auto parentheses
\newcommand{\lbra}[1]{\left[#1\right]} % Auto bracket
\newcommand{\eqskip}{\\[3 pt]}  % Spacing for equation
\newcommand{\eqskipp}{\\[4 pt]}  % Spacing for equation
\newcommand{\dt}{\mathrm{d}t}

\newcommand{\ddt}{\frac{\mathrm{d}}{\dt}}
\newcommand{\xboldot}{\dot{\mathbf{x}}}
\newcommand{\ubold}{\mathbf{u}}
\newcommand{\xbold}{\mathbf{x}}
\newcommand{\fbold}{\mathbf{f}}
\newcommand{\vm}{v_{\mathrm{m}}}
\newcommand{\vmdot}{\dot{v}_{\mathrm{m}}}

\newcommand{\xG}{x_{\mathrm{G}}}

\newcommand{\IzG}{I_{z\mathrm{G}}}
\newcommand{\Nm}{N_\mathrm{m}}
\newcommand{\Hsub}{_\mathrm{H}}

\newcommand{\optcontoff}{\mathbf{u}^{*}_{\mathrm{off}}}
\newcommand{\optstatoff}{\mathbf{x}^{*}_{\mathrm{off}}}
\newcommand{\onoptstate}{\mathbf{x}^{*}}
\newcommand{\onopcontrl}{\mathbf{u}^{*}}

\newcommand{\xfin}{\displaystyle{\xbold_{\mathrm{fin}}}}
\newcommand{\deltamax}{\displaystyle{\delta_{\mathrm{max}}}}
\newcommand{\nmax}{n_{\mathrm{max}}}
\newcommand{\sumstate}{\sum_{i=1}^{6}}

\newcommand{\hs}{h_\mathrm{s}}
\newcommand{\hd}{h_\mathrm{d}}

\newcommand{\actut}{\bar{t}}

\newcommand{\tfpla}{t_{\mathrm{f}}}
\newcommand{\tf}{\lpare{\tfpla}}

\newcommand{\tkpare}{\lpare{t_k}}

\newcommand{\ttwoqpare}{\lpare{t_{2q}}}
\newcommand{\actutjpare}{\lpare{\actut_j}}
\newcommand{\commeq}{\text{,}}
\newcommand{\Ns}{N_\mathrm{s}}
\newcommand{\Nc}{N_\mathrm{c}}
\newcommand{\dynamics}{\fbold\lpare{\xbold(t),\ubold(t), V, \chi}}
\newcommand{\dynaonl}{\fbold\lpare{\xbold(t),\ubold(t), V\actutjpare, \chi\actutjpare}}

\newcommand{\ttkclosed}{\lbra{t_k, t_{k+1}}}

\newcommand{\ttfclosed}{\lbra{0, \tfpla}}
\newcommand{\pbold}{\mathbf{p}}
\newcommand{\Pbold}{\mathbf{P}}
\newcommand{\bsub}{_{\mathrm{b}}}
\newcommand{\ssub}{_{\mathrm{s}}}
\newcommand{\bsubi}[1]{_{\mathrm{b},#1}}
\newcommand{\ssubi}[1]{_{\mathrm{s},#1}}
\newcommand{\nv}{n_{\mathrm{v}}}
\newcommand{\ns}{n_{\mathrm{s}}}
\newcommand{\uonint}{u_{\mathrm{off}}(0)}

\newcommand{\thetasi}{\theta_{\mathrm{s},i}}
\newcommand{\xopti}{x_{\mathrm{off},i}^{*}}
\newcommand{\tfopt}{t_{\mathrm{f,off}}}
\newcommand{\TWSCMA}{T_{\mathrm{w1}}}

\newcommand{\TNOWS}{T_{\mathrm{cs}}}
\newcommand{\qminsub}{_{2q-1}}
\newcommand{\qplussub}{_{2q+1}}

\newcommand{\MATLAB}{\textsc{Matlab}\xspace} % Matlab

\usepackage{ulem}

\newcommand\Erase{\bgroup\markoverwith{\textcolor{red}{\rule[.5ex]{2pt}{0.4pt}}}\ULon}
\def\tsc#1{\csdef{#1}{\textsc{\lowercase{#1}}\xspace}}
\tsc{WGM}
\tsc{QE}
\ExplSyntaxOn
\keys_set:nn{ stm / mktitle }{ nologo }
\ExplSyntaxOff

\begin{document}
\let\WriteBookmarks\relax
\def\floatpagepagefraction{1}
\def\textpagefraction{.001}

% Short title
\shorttitle{Warm-started Semionline Trajectory Planner for Automatic Docking}    
% Short author
\shortauthors{Dimas M. Rachman et~al.}  

% Main title of the paper
\title [mode = title]{Warm-started Semionline Trajectory Planner for Ship's Automatic Docking (Berthing)}  

% Title footnote mark
% eg: \tnotemark[1]
%\tnotemark[<tnote number>] 

% Title footnote 1.
% eg: \tnotetext[1]{Title footnote text}
%\tnotetext[<tnote number>]{<tnote text>} 

% Options: Use if required
% eg: \author[1,3]{Author Name}[type=editor,
%       style=chinese,
%       auid=000,
%       bioid=1,
%       prefix=Sir,
%       orcid=0000-0000-0000-0000,
%       facebook=<facebook id>,
%       twitter=<twitter id>,
%       linkedin=<linkedin id>,
%       gplus=<gplus id>]

% List of authors
\author[1]{Dimas M. Rachman}[]
\cormark[1]
\ead{dimas_rachman@naoe.eng.osaka-u.ac.jp}
\credit{Conceptualization, Methodology, Software, Writing - original draft}

\author[1]{Atsuo Maki}[]%[orcid=0000-0002-2819-1297]
%\ead{maki@naoe.eng.osaka-u.ac.jp}
\credit{Supervision, Writing - review \& editing}

\author[1]{Yoshiki Miyauchi}[]
\credit{Writing - review \& editing}

\author[1]{Naoya Umeda}[]
\credit{Supervision, Funding acquisition}

% Address/affiliation
\address[1]{Osaka University, 2-1 Yamadaoka, Suita, 565-0871, Osaka, Japan}

\cortext[cor1]{Corresponding author}

% Here goes the abstract
\begin{abstract}
In the usual framework of control, a reference trajectory is needed as the set point for a feedback controller. This reference trajectory can be generated by solving a trajectory optimization problem. This problem is a continuous optimal control problem (OCP) that is transcribed into a finite-dimensional nonlinear optimization problem (NLP) and solved by SQP. For an underactuated conventional vessel, the mathematical model can be very intricate, hence the NLP itself. This causes significant computational time. This article demonstrates that the balance between the feasibility of the reference trajectory and the computational time can be achieved for an underactuated vessel in a disturbed and restricted environment. This is done by: (1) using an almost-globally optimal offline solution as a warm start in a semionline trajectory optimization to speed up the calculation, (2) including the prediction of wind dynamics, and (3) representing the ship as a rigid body and using a predefined boundary to generate the necessary spatial constraints via a point-in-polygon method that ensure a collision-free trajectory in a nonconvex region. Incorporation of these three things maintains a safe and dynamically feasible trajectory where the warm start gives a considerable computational speedup and better results than that without a warm start.
\end{abstract}

% Use if graphical abstract is present
%\begin{graphicalabstract}
%\includegraphics{}
%\end{graphicalabstract}

% Research highlights
%\begin{highlights}
%\item An optimal-control-based semionline trajectory planner (SO-TP) for docking operation is demonstrated. 
%\item The SO-TP is warm-started by an almost-globally optimal solution from an offline trajectory planner to speed up the computational time.
%\item The warm-started SO-TP is applicable to underactuated vessels in any arbitrary shape of harbor area.
%\end{highlights}

% Keywords
% Each keyword is seperated by \sep

\maketitle
\sloppy

% Main text
\section{Introduction} \label{Sec:1-intro}
    % Show the general situation
    One of the most discussed matters in the problem of automatic docking is the trajectory planning of a docking operation. Usually, a trajectory is planned through \textit{trajectory optimization}, which is a case of continuous optimal control problem (OCP). Practically a continuous OCP is discretized and transcribed into a finite-dimensional nonlinear optimization problem (NLP) via \textit{direct} methods such as \textit{shooting} method and \textit{collocation} method. In this article, the term OCP is used interchangeably with trajectory optimization.
    
    % Difference between trajectory planning for berthing and general operation
    Trajectory planner for a docking operation is quite different from that for a general operation. The global nature of a general operation makes it possible for the trajectory to be obtained by smoothing a path in accordance to part of the ship's dynamics \citep{Vagale2021,ZHOU2020107043}. However, when planning a trajectory for a docking operation, the low speed and high safety requirement result in a more complicated dynamics and subsequently the trajectory optimization itself.
    
    % Important points of the previous development
    Research on a trajectory planner for docking operation sees various efforts to balance three things: low computational cost \citep{BITAR202014488,MARTISEN-DOCK}, applicability to underactuated conventional vessel \citep{Maki2020application,maki2020pt1,MIZUNO2015305}, and feasibility with respect to the surrounding dynamic \citep{HAN2021109352,MIZUNO2015305} and geometric information \citep{BITAR202014488,LIAO201947,MARTINSEN201997,miyauchi2021planner}. This is relevant, because in practice a shipmaster will try to replan a new trajectory for docking operation if it is considered not feasible anymore. Even an experienced shipmaster might require some time to plan a new trajectory. For this reason, to realize an autonomous ship, it is necessary to design a relatively fast planner that can replan a feasible and safe trajectory, and applicable to conventional vessels. 
    
    % Previous research on OCP-based trajectory planner
    This article limits the discussion within the scope of OCP-based trajectory planner that was pioneered by \cite{shouji1992automatic}. The trajectory planner introduced in \cite{Maki2020application} and \cite{miyauchi2021planner} can produce an almost-globally optimal, feasible, and safe trajectory for a conventional vessel. However, this planner requires a long computational time which makes it not possible for a real-time application. The one introduced in \cite{MIZUNO2015305} is relatively fast and also applicable to a conventional vessel, but there are no spatial constraints that can guarantee safe trajectory. On the other hand, \cite{MARTISEN-DOCK} showed a fast planner that guarantees a safe trajectory, but applicable to a fully/overly actuated vessel with a less intricate mathematical model. Still related to the discussion, \cite{HAN2021109352} presented a fast optimization-based planner for a conventional vessel, however, the ship is regarded just as a particle instead of a rigid body and the collision with the harbor's walls is avoided by expanding the harbor's boundaries.
    
    % Main intention, so the reader will get a glimpse of everything here
    By exploiting the advantages of the previously mentioned works, the main intention of this article is to introduce a \textit{semionline trajectory planner} (SO-TP) that is able to plan a collision-free trajectory for an underactuated vessel under wind disturbance. The term \textit{semionline} implies the nature of the computational time that is fast enough in practice, but not strictly in real time (online). This term also implies the ability of the planner to take into account the wind dynamics measured on the vessel when the planner is executed. 
    
    % One main point and several evaluation matters
    One of the key points in the SO-TP is to use the almost-globally optimal and feasible solution from \cite{Maki2020application} as a \textit{warm start}, i.e., good initial guess in the SO-TP. This warm start is expected to give a good starting point for the optimization inside the SO-TP so that the search for the local optimum can be done faster than that without a warm start (\textit{cold start}, using any arbitrary initial guess). To evaluate the performance of the warm-started SO-TP, the following matters are discussed: 
    \begin{enumerate}
        \item The applicability of just one warm start to speed up the SO-TP for various scenarios: different initial states and wind parameters.
        \item The comparison of the resulting trajectories with and without a warm start (linear initial guess). Results from a total of 14 different scenarios are compared based on the feasibility and the computational time.
        \item The comparison of the resulting trajectories with two different warm starts: (1) obtained by the method in \cite{Maki2020application}; (2) obtained by the SO-TP itself with a linear initial guess. This is to show the necessity for a good initial guess as the warm start.
    \end{enumerate}
    
    % Main improvements
    The readers will find this article as a continuation and implementation of the work in \cite{Maki2020application} and \cite{miyauchi2021planner} where computational time was the biggest issue that limits their practical use. To address this issue, the first major development in this article that is unique to the previous works on OCP-based docking trajectory planners for conventional vessels is the use of the warm start to speed up the computational time. 
    
    The next major development is that this article implements a collision constraint generator via a \textit{point-in-polygon} method \citep{HORMANN2001131} that is applicable to either convex or nonconvex region. Up to the limit of the authors' knowledge, this direct implementation is the first in the field of trajectory optimization for marine use. This can be an alternative to those introduced in \cite{HAN2021109352}, \cite{LIAO201947}, and \cite{MARTISEN-DOCK}. 
 
   \begin{figure*}[tbp]
        \centering
            \includegraphics[width=0.75\textwidth]{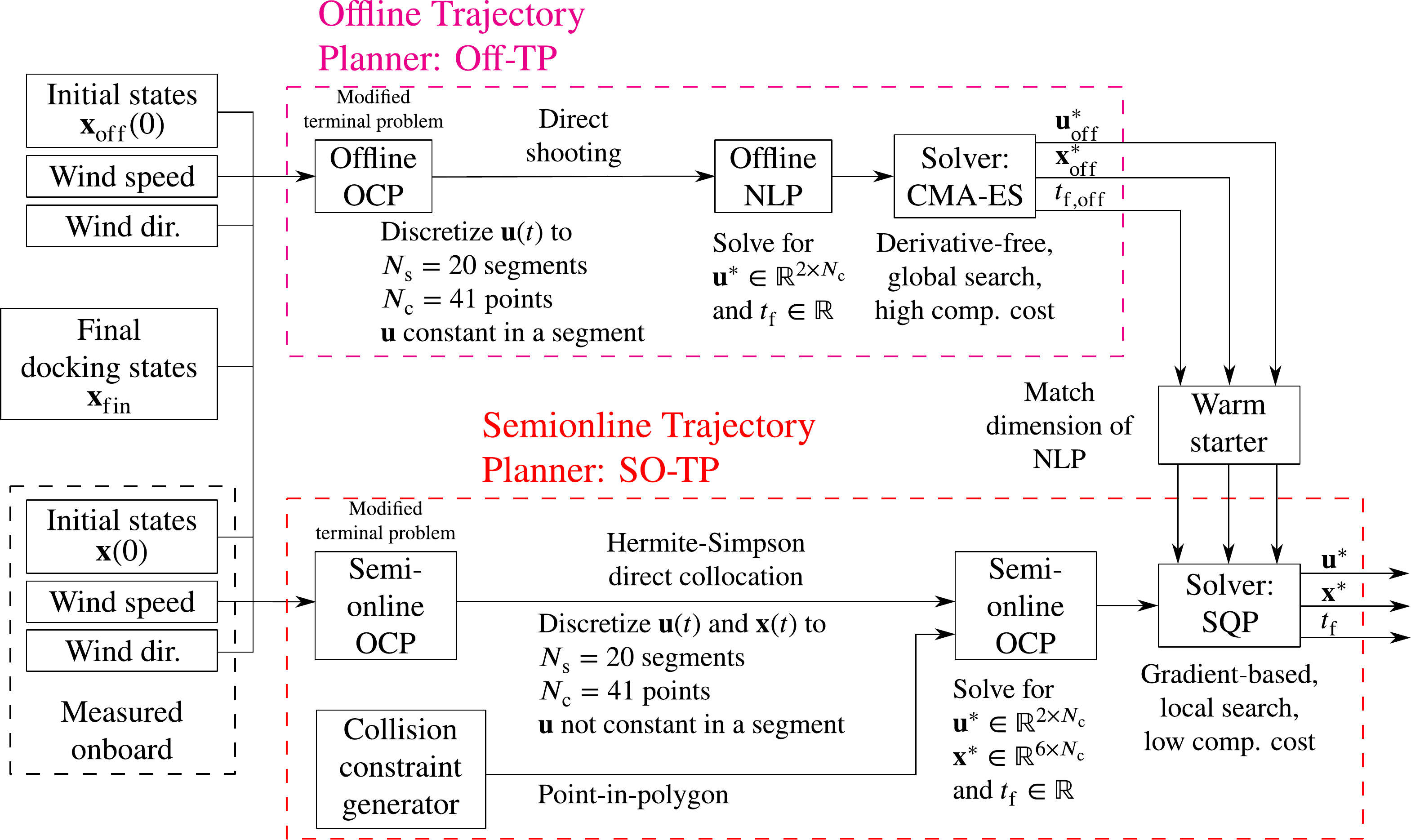}
            \caption{Outlines of the warm-started semionline trajectory planner.} 
            \label{fig:flowchart}
    \end{figure*}
    
    % Organization of the article
    For the remaining of the article, it is organized into another four sections. Section \ref{Sec:2-Proposed} explains the proposed scheme and the problem formulation: maneuvering model of the subject ship, brief explanation of the warm start, the SO-TP itself, and the collision constraints generator. Section \ref{Sec:3-SimulationConditions} and section \ref{Sec:4-SimRes} discuss the simulation condition and the results from several different cases, respectively. Finally, section \ref{Sec:5-ConcludeRemarks} concludes the article with additional remarks on several practical matters.

\section{Proposed Scheme and Formulation} \label{Sec:2-Proposed}
    The outlines of the proposed scheme is visualized in Fig. \ref{fig:flowchart} and can be summarized into the following steps:
    \begin{enumerate}
        \item First, optimal trajectories are generated. These optimal trajectories are termed \textit{offline trajectories} or \textit{offline solutions}, and the planner is called \textit{offline trajectory planner} (abbreviated as Off-TP). 
        \item The offline solutions can be obtained for several initial states and wind parameters. As a demonstration, this article shows how just one offline solution (without considering the wind) can be used in many scenarios.
        \item Then, the offline solution is used to warm-start the semionline trajectory planner (SO-TP).
        \item The OCP in the SO-TP is discretized and transcribed into a finite-dimensional NLP by direct Hermite-Simpson collocation method. The form of the objective function is identical to that in the Off-TP.
        \item With a \textit{point-in-polygon} method, one can check whether a point is inside or outside a polygon that defines an obstacle-free region. This results in several nonlinear and implicit spatial constraints to the NLP.
        \item Finally, the online NLP is solved by \MATLAB \lstinline{fmincon} package with SQP (\textit{sequential quadratic programming}) \citep{boggs_tolle_1995} as its main algorithm. An optimization where a good guess is provided is called a \textit{warm-started} optimization \citep{WarmStartIPForsgren}.
    \end{enumerate}
    In this article, the trajectory planner is simulated for an underactuated vessel: a 3 m \textit{Esso Osaka} 1/108 \textit{scale model} or just \textit{Esso Osaka} (the principal particulars are given in Table \ref{tab:A1-EssoOsakaPP} in the appendix). It is equipped with an electric motor that allows instantaneous mode switch and single rudder that can rotate at a rate of approximately $70$ degree/s.
    
    \subsection{Esso Osaka Maneuvering Model} \label{subSec:2.1-MMG}
        Normally a docking operation is carried out at a low surge velocity. For an underactuated conventional vessel: nonholonomic, it usually requires the dynamics for all possible combinations of surge velocity and propeller's modes to be described. For these reasons, the mathematical model of such a system can be very nonlinear and intricate.
        
        \begin{figure}[htbp]
            \centering
                \includegraphics[width=0.55\columnwidth]{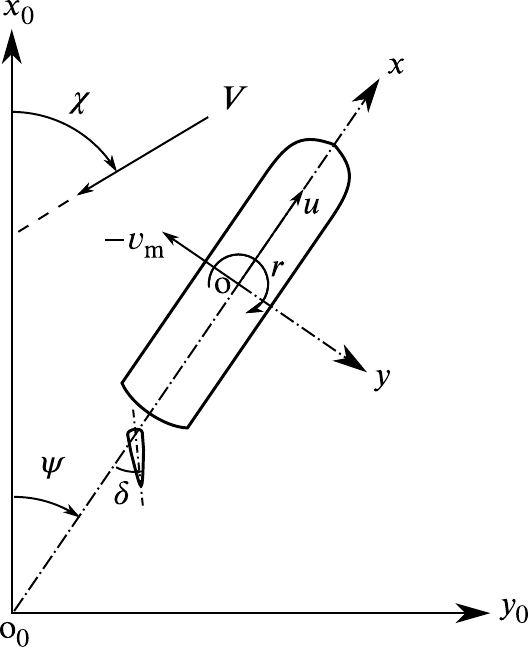}
                \caption{Coordinate systems.} 
                \label{fig:mmgcoordinate}
        \end{figure}       
        
        Two coordinate systems are introduced: a ship-fixed coordinate system $\mathrm{o}-xy$ where midship is its origin and an earth-fixed coordinate system $\mathrm{o}_0-x_0y_0$. Angle between the two coordinate systems is the ship's yaw angle $\psi$ and $r$ denotes its angular velocity. The surge and sway velocities at the midship are denoted by $u$ and $\vm$, respectively.
        
        The kinematics at the midship can be transformed from the ship-fixed coordinate system to the earth-fixed coordinate system by the following transformation,
        \begin{align}\label{eq:coordtransform}
            \ddt
            \begin{bmatrix}
                x_0 \eqskip y_0 \eqskip \psi
            \end{bmatrix} =
            \begin{bmatrix}
                \cos{\psi} & -\sin{\psi} & 0 \eqskip
                \sin{\psi} & \cos{\psi} & 0 \eqskip
                0 & 0 & 1
            \end{bmatrix}
            \begin{bmatrix}
                u \eqskip \vm \eqskip r
            \end{bmatrix}.
        \end{align}
            
        The kinetics of Esso Osaka is based on the standard equations of motion presented by the MMG (maneuvering model group) \citep{mmgmodel} as follows:
        \begin{subequations}\label{equofmotion}
            \begin{align}
                \lpare{m+m_x}\dot{u}-\lpare{m+m_y}\vm r-\xG mr^2 &= X  \eqskip
                \lpare{m+m_y}\vmdot+\lpare{m+m_x}ur+\xG m\dot{r} &= Y \eqskip
                \lpare{\IzG+\xG^2m+J_z}\dot{r}+\xG m\lpare{\vmdot+ur} &= \Nm \commeq 
            \end{align}
        \end{subequations}
        and can be rearranged into,
        \begin{subequations}\label{eq:rearrangedeqofmotion}
        \begin{align}
            \dot{u} &=
            \frac{X + M_y \vm r + \xG m r^2}{M_x} \eqskipp
            \vmdot &=
            \frac{\lpare{Y - M_x u r}I_{z\mathrm{m}} - \lpare{\Nm - \xG m u r }\xG m}{M_y I_{z\mathrm{m}}-\lpare{\xG m}^2} \eqskipp
            \dot{r} &=
            \frac{\lpare{Y-M_x ur}\xG m - \lpare{\Nm - \xG m u r}M_y}{ \lpare{\xG m}^2 - M_y I_{z\mathrm{m}}}\commeq
        \end{align}
        \end{subequations}
        where simplified terms: $M_x = m + m_x$, $M_y = m + m_y$, and $I_{z\mathrm{m}}=\IzG+x^2_{\mathrm{G}} m + J_z$ are used. Here, $m$ and $\IzG$ denote the ship's mass and moment of inertia about center of gravity $\xG$, respectively. Moreover, $m_x$, $m_y$, and $J_z$ denote the added mass and moment of inertia in each respective axis. 
            
        $X$ and $Y$ are the total surge and sway force, respectively, while $\Nm$ is the total moment about the midship. These total forces are contributions from hull forces \citep{yoshimuralow}, four-quadrant steering and thrust forces \citep{Kitagawa2015,Miyauchi2021SI,mmgmodel}, and wind forces \citep{fujiwind}. The important coefficients are given in \cite{maki2020pt1} and \cite{Miyauchi2021SI}.
            
        Combining the kinematics (\ref{eq:coordtransform}) and the kinetics, (\ref{eq:rearrangedeqofmotion}) one can obtain the dynamics that can be rearranged into a simplified expression,
        \begin{align}\label{eq:simpledyna}
            \xboldot(t) = \dynamics\commeq
        \end{align}
        where $V$ is the true wind speed that \textit{blows-from} at an angle $\chi$ from the North (the positive $x_0$ axis), and
        \begin{align}
            \xbold(t) &\triangleq \begin{bmatrix}
                        \; x_0 \;\;\; u \;\;\; y_0 \;\;\; \vm \;\;\; \psi \;\;\; r \;\,
                     \end{bmatrix}^\top \label{statevect} \eqskip
            \ubold(t) &\triangleq \begin{bmatrix}
                        \; \delta \;\;\; n \;\,
                     \end{bmatrix}^\top \label{controlvect}.    
        \end{align}
        $\xbold(t) \in \mathbb{R}^6$ is termed \textit{states} and $\ubold(t) \in \mathbb{R}^2$ is termed \textit{control}, where $\delta$ is rudder angle in degree and $n$ is propeller revolution number in rps (1/s).
        
    \subsection{Offline Trajectory Planner: Off-TP}
        \label{subSec:2.2-OffTrajOpt}
        The Off-TP is an OCP-based trajectory planner which role is to provide a \textit{warm start}, i.e., an almost-globally optimal solution \citep{WarmStartIPForsgren}, as the initial guess to the SO-TP so that it can converge faster. This warm start does not necessarily need to be the solution to the exact same optimization problem.
        
        In the Off-TP, the OCP is discretized and transcribed into a finite-dimensional NLP by shooting method. The goal is to find the optimal control $\optcontoff$ at each discretization point that drive the system from the initial states $\xbold(0)$ to the desired final docking states $\xfin$. This optimal control gives optimal state trajectories $\optstatoff$ that minimize a modified terminal objective function: minimize the deviation of the states from $\xfin$ at the final time $\tfopt$. These offline solutions from the Off-TP: $\optcontoff$, $\optstatoff$, and $\tfopt$ are used as a warm start for the SO-TP (see Fig. $\ref{fig:flowchart}$). Note that the terms with asterisk ($^*$) denote the corresponding optimal values. 
        
        The offline NLP is solved by a nonconvex and derivative-free global optimization technique known as the evolution strategy with covariance matrix adaptation (CMA-ES) \citep{hansencma2003,auger2005restart,sakamoto2017modified}\@. Despite being computationally demanding (a stochastic technique, coupled with explicit integration of the dynamics and single shooting method), it performed very well to locate the best approximation of the global optimum. So, the offline solution from the Off-TP can be considered almost-globally optimal. Details about the Off-TP are explained in \cite{Maki2020application}.        
        
        \subsection{Semionline Trajectory Planner: SO-TP} \label{subSec:2.2-OnTrajOpt}
        The SO-TP is developed upon the idea that in practice there may be a need to \textit{replan} the trajectory based on the surrounding conditions. It is desired that the updated trajectory is kept feasible and safe.
        
        Let $\actut_j$ be the $j$-th sampling of the actual time $\actut$ outside the SO-TP. Let $t$ be the normalized prediction time inside the SO-TP such that $t=0$ is $\actut_j$. Then, the continuous (a modified terminal OCP, before transcription) semionline trajectory optimization problem at $\actut_j$ can be stated as:
        \begin{subequations}\label{eq:onobj}
        \begin{flalign}
        &\text{determine} && \delta^{*}(t) \commeq \;  n^{*}(t) \commeq \; \tfpla \nonumber \eqskip
        &\text{that min.}&& J_{\mathrm{on}}= \sumstate \lpare{x_{i}\lpare{\tfpla}-x_{\mathrm{fin}, i}}^2 \nonumber \eqskip
        & && \hphantom{J_{\mathrm{on}}=} \times \int_{0}^{\tfpla} \sumstate \lpare{x_{i}\lpare{t}-x_{\mathrm{fin}, i}}^2 \, \dt \label{eq:Jon} \eqskip 
        &\text{subject to} && \xboldot(t) = \dynaonl \commeq \label{eq:ondynocp} \eqskip
        & &&\xbold^*\lpare{0}=\xbold\actutjpare \commeq \; \xbold^*\tf=\xfin \commeq \eqskip
        & &&-\deltamax \leq \delta(t) \leq \deltamax \commeq \label{eq:ocprudconst} \eqskip
        & &&-\nmax \leq n(t) \leq \nmax \commeq \label{eq:ocppropconst} \eqskip
        & && t\in \ttfclosed \commeq \;\; \text{ and } \;\; \tfpla \in \left(0,\infty \right). \nonumber
        \end{flalign} 
        \end{subequations}
        The wind parameters are assumed constant when planning the trajectory, i.e., $V(t)=V\actutjpare$ and $\chi(t)=\chi\actutjpare$ for all $t$.
        
        The objective function of the online OCP $J_{\mathrm{on}}$ (\ref{eq:Jon}) will drive the optimization to minimize the deviation of the final states $\xbold\tf$ from the desired final docking states $\xfin$. The first summation term in (\ref{eq:Jon}) is the deviation at $\tfpla$ and the integral term captures the accumulation of the deviation of the states $\xbold(t)$ from $\xfin$ within the whole time. The integral term will also indirectly force the optimization to minimize the final time $\tfpla$. This objective function is identical to that in the Off-TP because ideally the objective of the Off-TP and the SO-TP should be kept as identical as possible.
        
        It is practical to plan a reference trajectory that is "easy" to be tracked, i.e., with less control effort. This is to compensate for the uncertainties that may affect the feedback controller. It can be done either by: adding control cost in the objective function (\ref{eq:Jon}) or explicitly limiting the allowable control (to $\deltamax$ and $\nmax$, less than the actual capacity). The latter is preferable since it is binding in its box constraint form: (\ref{eq:ocprudconst}) and (\ref{eq:ocppropconst}), see Table \ref{tab:1-controlmaxreason}.
        
        \begin{table}[pos=h]
        \caption{Control limitations in generating the reference trajectory.}\label{tab:1-controlmaxreason}
            \begin{tabular*}{\tblwidth}{@{}CCC@{}}
                \toprule
                 Control & Limiting values & Actual capacity \\ 
                \midrule
                $\delta$ (deg) & $\deltamax = 25$ & $35$ \\
                $n$ (rps) & $\nmax = 15$ & $20$ \\
                \bottomrule
            \end{tabular*}
        \end{table} 
        
        \subsubsection{Separated Hermite-Simpson Collocation} \label{subsubSec:2.3.1-Herm-Sims-Coll}
            The continuous OCP in (\ref{eq:onobj}) can be discretized and transcribed into a finite-dimensional nonlinear optimization problem (NLP) by direct collocation. The dynamics, the state, and the control trajectories are discretized into $\Ns\in\mathbb{N}$ equal-spaced segments; each bounded by \textit{knot} points. The segment's length (between two knot points) is $\hs=\tfpla/\Ns$.
            
            In \textit{separated} Hermite-Simpson collocation method \citep{doi:BettsControl}, a \textit{collocation} point at the mid-segment is required so that the length between each discretization point is $\hd=0.5\hs$. Along with the knot points, the number of \textit{discretization} points is $\Nc=2\Ns+1$.
            
            Now, let $q$ denote the $q$-th segment such that $q=1,2,\dots,\Ns$ and let $k$ denote the $k$-th discretization point such that $k=1,2, \dots, \Nc$. With these notations, odd $k$ is the knot point where $k=2q-1$ and even $k$ is the collocation point where $k=2q$. Note that these two representations of the discretization points (by $k$ and $q$) are equivalent and will be used interchangeably.
            
            Moreover, let $t_k$ be the time at the $k$-th discretization point. For brevity, $\xbold_k\equiv\xbold\tkpare$ and $\ubold_k\equiv\ubold\tkpare$ denote the optimization variables, i.e., the states (\ref{statevect}) and the control (\ref{controlvect}) at the discretization point. Following the same notation, $\fbold_k\equiv\fbold\lpare{\xbold_k,\ubold_k,V,\chi}$ denotes the true dynamics (\ref{eq:simpledyna}) evaluated at the discretization point.
            
            \begin{figure}[htbp]
            \centering
                \includegraphics[width=0.95\columnwidth]{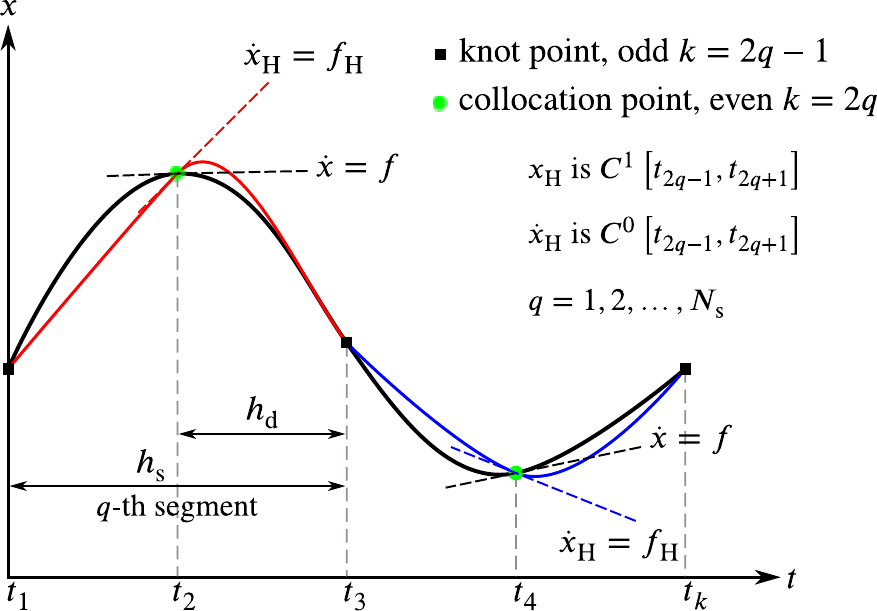}
                \caption{Hermite-Simpson collocation: match the slope of the interpolant $\fbold\Hsub$ with the true dynamics $\fbold$ at the collocation point.} 
                \label{fig:HS}
            \end{figure}              
            
            The idea of Hermite-Simpson collocation is to approximate the state trajectory $\xbold(t)$ within a segment (shown as black curves in Fig. \ref{fig:HS}) by first-order third-degree (cubic) \textit{Hermite interpolation polynomial}, denoted by $\xbold\Hsub(t)$, with $C^{1}$ continuity (shown as red and blue curves in Fig. \ref{fig:HS}). These piecewise polynomial functions form a continuous cubic Hermite spline that satisfies both the states and its first derivative (true dynamics) at all discretization points. On the other hand, the control trajectory in a segment can be approximated by a quadratic, linear, or constant function. This function is denoted by $\ubold\Hsub(t)$.
            
            The key point of Hermite-Simpson collocation method is to look for the states and control such that at the collocation point, the slope of the Hermite cubic interpolant, i.e., $\xboldot\Hsub$ (or $\fbold\Hsub$), is equal to the true dynamics $\fbold$. This is visualized in Fig. \ref{fig:HS} where the red dashed line should match the black dashed line. To do this, first the states at the collocation point can be interpolated by,
            \begin{align}\label{eq:statinterp}
                \begin{split}
                    \xbold\Hsub\ttwoqpare&=\frac{1}{2}\lpare{\xbold_{2q-1}+\xbold\qplussub}\eqskip
                    &\hphantom{=}+\frac{\hs}{8}\lpare{\fbold\qminsub-\fbold\qplussub}\commeq
                \end{split}
            \end{align}
            which results in additional constraints as follows,
            \begin{align}
                \xbold\Hsub\ttwoqpare&=\xbold_{2q} \label{eq:interpconstraintHS1}\text{.}
            \end{align}
            Similarly, the dynamics can be interpolated by,
            \begin{align}\label{eq:dynainterp}
                \begin{split}
                    \xboldot\Hsub\ttwoqpare&=\fbold\Hsub\ttwoqpare\eqskip
                    &=-\frac{3}{2\hs}\lpare{\xbold\qminsub-\xbold\qplussub}\eqskip
                    &\;\hphantom{=}-\frac{1}{4}\lpare{\fbold\qminsub+\fbold\qplussub}\text{.}\eqskip
                \end{split}
            \end{align}    
            With $\xbold_{2q}$ that satisfies (\ref{eq:interpconstraintHS1}) and the control $\ubold_{2q}$, one can directly calculate the true dynamics at the collocation point $\fbold_{2q}$ from (\ref{eq:simpledyna}) and equating it with the interpolated dynamics (\ref{eq:dynainterp}). This results in additional constraints as follows,
            \begin{align}
                \fbold\Hsub\ttwoqpare&=\fbold_{2q}\commeq\label{eq:quadconstraintHS1}
            \end{align}
            which, if satisfied, is equivalent to implicitly integrating the true dynamics within a segment by Simpson quadrature rule:
            \begin{align*}
                \xbold\qplussub - \xbold\qminsub \approx \frac{\hs}{6}\lpare{\fbold\qminsub+4\fbold_{2q}+\fbold\qplussub}\text{.}
            \end{align*}
            
            In summary, at each segment $q$ there are two additional constraints for the NLP, i.e., the \textit{Hermite interpolation constraint} (\ref{eq:interpconstraintHS1}) and the \textit{Simpson quadrature constraint} (\ref{eq:quadconstraintHS1}); both imposed at the collocation points. The goal of the NLP is to find the final time along with the control and the states at all discretization points that satisfy all the constraints.\\
        
        \subsubsection{Point-in-Polygon: Collision Constraints}
            \label{subsubSec:2.3.2-PIPColConstr} 
            Suppose $\Pbold\bsub \in\mathbb{R}^{2\times\nv}$ is a matrix whose columns are the vertices position in the earth-fixed coordinate system,
            \begin{align*}
                \Pbold\bsub \triangleq
                    \begin{bmatrix} \; \pbold\bsubi{1} \;\;\; \pbold\bsubi{2} \;\;\; \dots \;\;\; \pbold\bsubi{\nv} \end{bmatrix}\commeq 
            \end{align*}
            where $\pbold\bsub=\lbra{\;x\bsub\;\;y\bsub\;}^{\top}$. The number of the vertices, $\nv\geq3$ is necessary to form a simple polygon: a closed region.
            
            \begin{figure}[htbp]
            \centering
                \includegraphics[width=0.64\columnwidth]{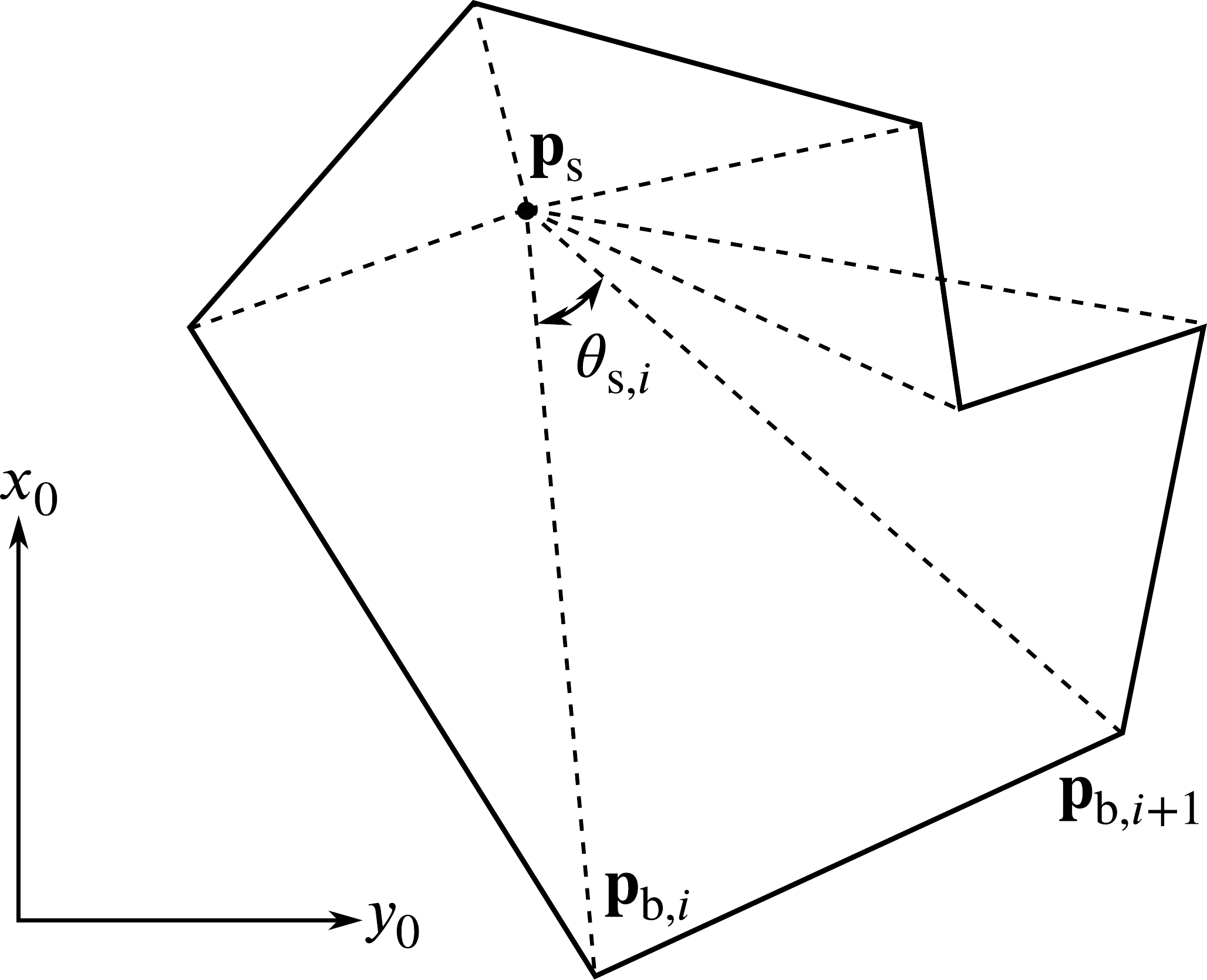}
                \caption{Sum of all $\thetasi$ (for $i=1,\dots,\nv$) must be equal to $2\pi$ for point $\pbold\ssub$ to be inside the polygon.} 
                \label{fig:sumofangle}
            \end{figure}            
            
            Let $\pbold\ssub=\lbra{\;x\ssub\;\;y\ssub\;}^{\top}$ be the point to be checked. If $\pbold\ssub$ is inside the polygon formed by $\Pbold\bsub$, then the sum of angle between all pairs of lines formed by $\pbold\ssub$ and two adjacent polygon vertices, i.e., $\pbold\bsubi{i}$ and $\pbold\bsubi{i+1}$, must be equal to $2\pi$. If this is not true, then $\pbold\ssub$ is outside the polygon. From Fig. \ref{fig:sumofangle}, let $\thetasi$ be the angle between two adjacent lines. 
        
            Suppose now $\Pbold\ssub \in\mathbb{R}^{2\times\ns}$ is a matrix whose columns are the points of the ship's boundary in the earth-fixed coordinate system,
            \begin{align*}
                \Pbold\ssub \triangleq
                    \begin{bmatrix} \; \pbold\ssubi{1} \;\;\; \pbold\ssubi{2} \;\;\; \dots \;\;\; \pbold\ssubi{\ns} \end{bmatrix} \commeq 
            \end{align*}
            where $\ns$ is the number of the ship's boundary points (not to be confused with the discretization points in the collocation). The boundary points represent the ship as a rigid body instead of a particle; ensures stringent constraints. These points depend on the position of midship $\lpare{x_0, y_0}$ and the yaw angle $\lpare{\psi}$, i.e., some of the optimization variables. Due to this dependency and the requirement that the sum of angle is equal to $2\pi$, one can construct additional equality constraints for the NLP as follows:
            \begin{align}\begin{split}
                \left.\sum_{i=1}^{\nv}\theta\ssubi{i}\right\rvert_{j,k} = 2\pi. \label{eq:sumofangle}
            \end{split}\end{align}
            This means that at each discretization point $k=1,2,\dots,\Nc$, (\ref{eq:sumofangle}) must be evaluated for all ship's boundary points $j=1,2,\dots,\ns$, which results in $\ns\times\Nc$ equalities.
            
            This method is equivalent to \textit{winding number }method \citep{HORMANN2001131} and known as \textit{angle summation} method. In CAD, it is not considered the best method to check whether a point is inside or outside a polygon. Nevertheless, in the scope of this article, it can be regarded as one of the most practical: analytically differentiable and can handle any shape of the harbor area.
            
        \subsubsection{Transcribed OCP in the SO-TP}
            \label{subsubSec:2.3.3-TranscribedOCP} 
            Based upon the previous two sub-subsections, the semionline OCP (\ref{eq:onobj}) can be transcribed to the following finite-dimensional NLP:
            \begin{subequations}\label{eq:transOCP}
            \begin{flalign}
            &\text{determine} \quad&&\delta^{*}_1, \delta^{*}_2, \dots, \delta^{*}_{\Nc} \in \mathbb{R} \commeq \nonumber \eqskip  
            & \quad && n^{*}_1, n^{*}_2, \dots, n^{*}_{\Nc} \in \mathbb{R} \commeq \nonumber \eqskip
            & \quad && \onoptstate_1, \onoptstate_2, \dots, \onoptstate_{\Nc} \in \mathbb{R}^6 \commeq \;\; \tfpla \in \mathbb{R} \nonumber \eqskip
            &\text{that min.} \quad&& J_{\mathrm{on}}= \sumstate \lpare{x_{i}\lpare{\tfpla}-x_{\mathrm{fin}, i}}^2 \nonumber \eqskip
            & \quad&& \hphantom{J_{\mathrm{on}}=} \times \int_{0}^{\tfpla} \sumstate \lpare{x_{i}\lpare{t}-x_{\mathrm{fin}, i}}^2 \, \dt \label{eq:transobj} \eqskip 
            &\text{subject to} \quad&&
            \onoptstate_1 = \xbold\actutjpare \commeq \; \onoptstate_{\Nc}=\xfin \commeq \label{eq:initstaton} \eqskip
            & \quad && \xbold_{\Nc-2}= \xfin \,- \nonumber \eqskip
            & \quad && \hphantom{\xbold_{\Nc-2} } \frac{\hs}{6}\lpare{\fbold_{\Nc-2} + 4\fbold_{\Nc-1} + \fbold_{\Nc}} \commeq \, \label{eq:finstaton} \eqskip
            & \quad && \xbold\Hsub\ttwoqpare = \xbold_{2q} \commeq \;\;\, q=1,2,\dots,\Ns \commeq \label{eq:intpconston} \eqskip
            & \quad && \fbold\Hsub\ttwoqpare= \fbold_{2q} \commeq \quad q=1,2,\dots,\Ns \commeq \label{eq:quadconston} \eqskip
            & \quad && \left.\sum_{i=1}^{\nv}\theta\ssubi{i}\right\rvert_{j,k} = 2\pi \commeq\quad j=1,2,\dots,\ns \label{eq:sumofangconst}\commeq\eqskip
            & \quad && -\deltamax \leq \delta\tkpare \leq \deltamax \commeq \label{eq:rudconstrainton} \eqskip
            & \quad && -\nmax \leq n\tkpare \leq \nmax \commeq \text{ and }  \label{eq:propconstrainton} \eqskip
            & \quad && \tfpla \; \in\, \left(0,\infty \right).  \nonumber
            \end{flalign} 
            \end{subequations}
            The integral term in the objective function (\ref{eq:transobj}) is approximated by Simpson quadrature.
    
            The goal of the NLP (\ref{eq:transOCP}) is to find $\lpare{2+6}\Nc+1$ number of optimal variables that minimize the objective function (\ref{eq:transobj}). The $6\times 2$ equalities in (\ref{eq:initstaton}) and (\ref{eq:finstaton}) enforce the initial and final states constraints, while (\ref{eq:intpconston}) and (\ref{eq:quadconston}) are the $6\times\Ns$ interpolation constraints and $6\times\Ns$ quadrature constraints, respectively. The $\ns\times\Nc$ equalities in (\ref{eq:sumofangconst}) are the collision constraints and the last two box constraints: (\ref{eq:rudconstrainton}) and (\ref{eq:propconstrainton}) saturate the control to the given values. 
            
            Sequential quadratic programming (SQP) from \MATLAB \lstinline{fmincon} package is used to solve the NLP (\ref{eq:transOCP}). It is a Newton's method with quadratic local convergence rate \citep{boggs_tolle_1995}\@. With a good initial guess (warm start), less iterations are expected, resulting in a faster convergence to the local optimum.

            The next chapter shows how just one offline solution from the Off-TP $\lpare{\optstatoff,\optcontoff,\tfopt}$ for a no-wind condition can be used to warm-start the SO-TP in various situations.\\ 
            
\section{Simulation Conditions}
    \label{Sec:3-SimulationConditions}
    The offline solutions $\optstatoff$ and $\optcontoff$ are obtained by the Off-TP with $V=0$ and $\chi=0$ (no wind). The initial states $\optstatoff(0)$, the desired final states $\xfin$, the final states $\optstatoff\lpare{\tfopt}$, and the final time $\tfopt$ of this offline solution are as follows:
    {\setlength{\mathindent}{0pt}
    \begin{align}
        \optstatoff(0) &= \lbra{\, 16.50 \;\; 0.12 \;\, -7.50\;\; 0.00\;\; 2\pi/3 \;\; 0.00\,}^{\top}\nonumber\eqskip
        \xfin &= \lbra{\, -0.50 \;\; 0.01 \;\, -0.50\;\; 0.00\;\; \pi\;\; 0.00\,}^{\top}\nonumber\eqskip
        \optstatoff\lpare{\tfopt}&= \lbra{\, -0.51 \;\; 0.01 \;\, -0.51\;\; 0.00\;\; 3.14\;\; 0.00\,}^{\top}\nonumber\eqskip
        \tfopt &= 160 \text{ s}\nonumber.
    \end{align}}
    
    At the final docking position, the ship's target berthing velocity (velocity normal to the wall, i.e., sway velocity $\vm$ if the ship is parallel to the berth's wall) is between 0.03 m/s to 0.04 m/s as reported in a statistical study of berthing velocity in the Port of Rotterdam \citep{Roubos2017}. The longitudinal velocity ($u$) on the other hand, should be kept as low as possible, but does not necessarily need to be zero as the port mooring/rigging will be used to completely fix the ship. So, the target surge velocity is set to 0.01 m/s (equivalent to 0.1 m/s in the full-scale ship) to relax the overall optimization problem.
    
    The performance of the warm-started SO-TP is evaluated for four main cases (M1 to M4) and ten additional arbitrary trial cases (A1 to A10, see Table \ref{tab:2-CaseInit}). The initial states for each cases are expressed as $x_i\actutjpare=d_i\xopti(0)$ for $i=1,\dots,6$, where $d_i$ is the element of a vector of multipliers $\mathbf{d}$. To quantify "how different" the initial states are from $\optstatoff(0)$, the Euclidean norm is introduced as, 
    \begin{align}
        L=&\sqrt{\sum_i^{6}\lbra{{\lpare{x_i\actutjpare-\xopti(0)}/\xopti}(0)}^2}\nonumber\\
        =&\sqrt{\sum_i^{6}\lpare{d_i - 1}^2}.
    \end{align}
    
    \begin{table*}\centering
        \caption{Evaluation cases for the SO-TP, see Fig. \ref{fig:App1} to Fig. \ref{fig:App3} in the appendix.}
        \label{tab:2-CaseInit}       
            \begin{tabular}{cccccccc}
                \hline\noalign{\smallskip}
                \multirow{2}{*}{Case} & Multiplier $\mathbf{d}$ & \multicolumn{2}{c}{Initial velocity $u(0)$} & \multirow{2}{*}{$\psi(0)$} & $V\actutjpare$ & $\chi\actutjpare$ & Norm $L=$\\
                \noalign{\smallskip}\cline{3-4}\noalign{\smallskip}
                    
                & $x_i\actutjpare=d_i\xopti(0)$ & (m/s) & Full scale (knots) & & (m/s) & (deg) & $\sqrt{\sum_i^{6}\lpare{d_i - 1}^2}$\\
                    
                \noalign{\smallskip}\hline\noalign{\smallskip}
                    
                M1 & $\lbra{\, 1.0 \;\; 1.0 \;\; 1.0 \;\; 1.0 \;\; 1.0\;\; 1.0 \,}^\top$ & $0.12$ & $2.43$ & $2\pi/3$ & \multicolumn{2}{c}{No wind} & $0.00$ \\
                    
                M2 & $\lbra{\, 1.1 \;\; 2.0 \;\; 1.1 \;\; 1.0 \;\; 1.2\;\; 1.0 \,}^\top$ & $0.24$ & $4.86$ & $4\pi/5$ & $0.75$ & $45$ & $1.03$ \\
                    
                M3 & $\lbra{\, 0.9 \;\; 2.0 \;\; 0.8 \;\; 1.0 \;\; 0.9\;\; 1.0 \,}^\top$ & $0.24$ & $4.86$ & $3\pi/5$ & $0.50$ & $225$ & $1.03$ \\        
                    
                M4 & $\lbra{\, 1.0 \;\; 3.0 \;\; 0.9 \;\; 1.0 \;\; 1.5\;\; 1.0 \,}^\top$ & $0.36$ & $7.28$ & $\pi$ & $0.75$ & $45$ & $2.06$ \\ 
                
                A1 & $\lbra{\, 1.2 \;\; 1.2 \;\; 1.1 \;\; 1.0 \;\; 1.1\;\; 1.0 \,}^\top$ & $0.14$ & $2.83$ & $11\pi/15$ & $0.75$ & $45$ & $0.32$  \\   
                
                A2 & $\lbra{\, 1.2 \;\; 1.4 \;\; 1.2 \;\; 1.0 \;\; 1.2\;\; 1.0 \,}^\top$ & $0.17$ & $3.44$ & $4\pi/5$ & $0.75$ & $180$ & $0.53$  \\
                
                A3 & $\lbra{\, 1.3 \;\; 1.6 \;\; 0.9 \;\; 1.0 \;\; 1.4\;\; 1.0 \,}^\top$ & $0.19$ & $3.84$ & $14\pi/15$ & $0.75$ & $125$ & $0.79$  \\  
                
                A4 & $\lbra{\, 1.4 \;\; 1.8 \;\; 0.8 \;\; 1.0 \;\; 0.9\;\; 1.0 \,}^\top$ & $0.22$ & $4.45$ & $3\pi/5$ & $0.50$ & $135$ & $0.92$  \\ 
                
                A5 & $\lbra{\, 1.5 \;\; 0.8 \;\; 1.2 \;\; 1.0 \;\; 1.1\;\; 1.0 \,}^\top$ & $0.10$ & $2.02$ & $11\pi/15$ & $0.50$ & $90$ & $0.58$  \\    
                
                A6 & $\lbra{\, 1.0 \;\; 0.8 \;\; 0.7 \;\; 1.0 \;\; 1.2\;\; 1.0 \,}^\top$ & $0.10$ & $2.02$ & $4\pi/5$ & $0.50$ & $315$ & $0.41$  \\
                
                A7 & $\lbra{\, 1.2 \;\; 2.2 \;\; 0.8 \;\; 1.0 \;\; 0.8\;\; 1.0 \,}^\top$ & $0.26$ & $5.26$ & $8\pi/15$ & $0.50$ & $250$ & $1.25$  \\ 
                
                A8 & $\lbra{\, 1.2 \;\; 2.0 \;\; 0.0 \;\; 1.0 \;\; 1.5\;\; 1.0 \,}^\top$ & $0.24$ & $4.86$ & $\pi$ & $0.50$ & $90$ & $1.51$  \\  
                
                A9 & $\lbra{\, 1.0 \;\; 1.0 \;\; -1.0 \;\; 1.0 \;\; 2.0\;\; 1.0 \,}^\top$ & $0.12$ & $2.43$ & $4\pi/3$ & $0.75$ & $0$ & $2.24$  \\   
                
                A10 & $\lbra{\, 0.8 \;\; 1.5 \;\; -0.9 \;\; 1.0 \;\; 2.2\;\; 1.0 \,}^\top$ & $0.18$ & $3.64$ & $22\pi/15$ & $0.50$ & $45$ & $2.31$  \\
                \noalign{\smallskip}\hline
            \end{tabular}
        \end{table*}            
            
    \subsection{Spatial/Collision Constraints Definition}\label{subSec:3.1-SpatConstDef}
\begin{figure}[htbp]
            \begin{subfigure}[b]{0.48\columnwidth}
            \centering
                \includegraphics[width=0.95\columnwidth]{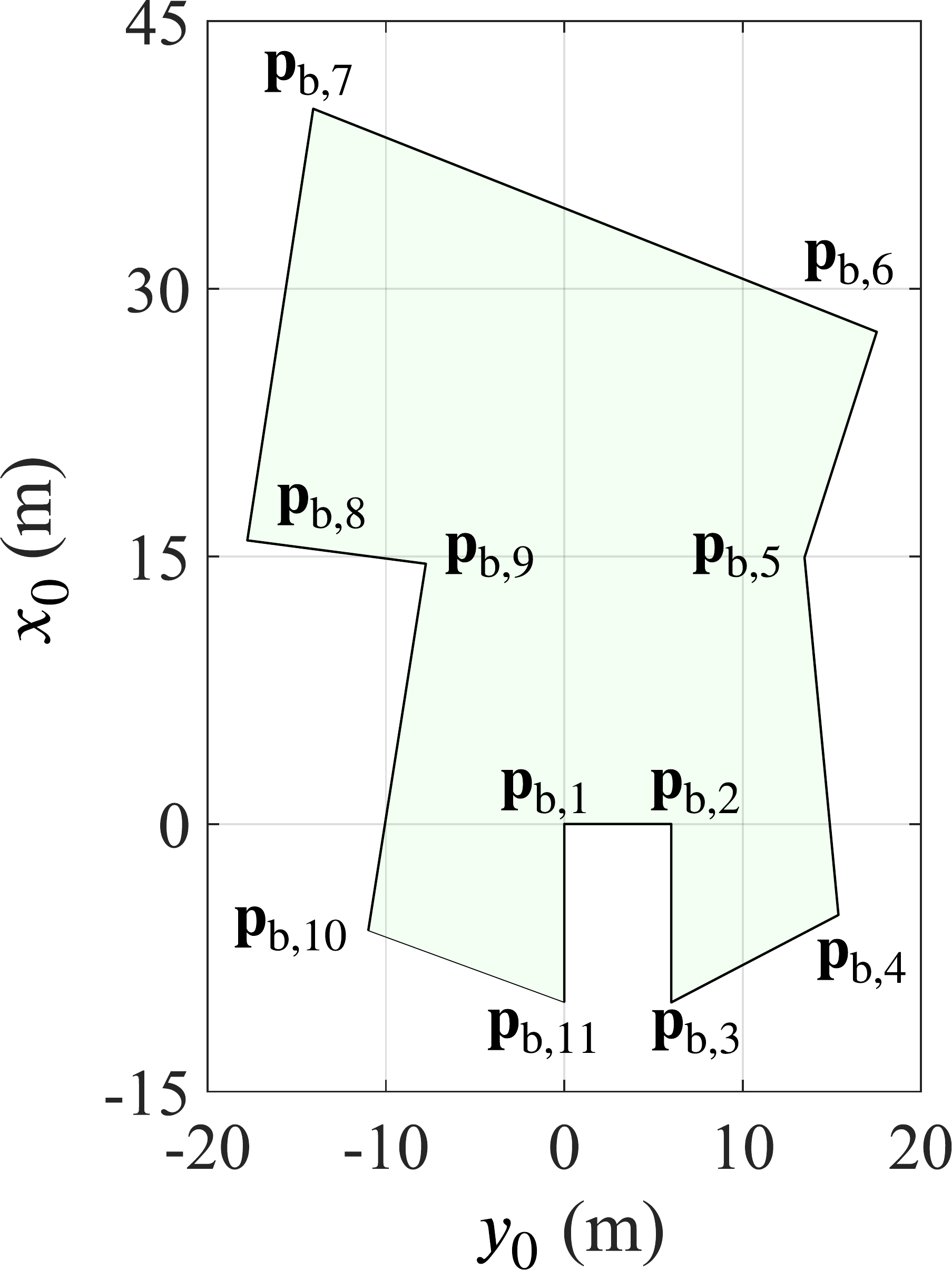}
                \caption{\label{fig:fig:boundary}}
            \end{subfigure}
            \begin{subfigure}[b]{0.48\columnwidth}
            \centering
                \includegraphics[width=0.5\columnwidth]{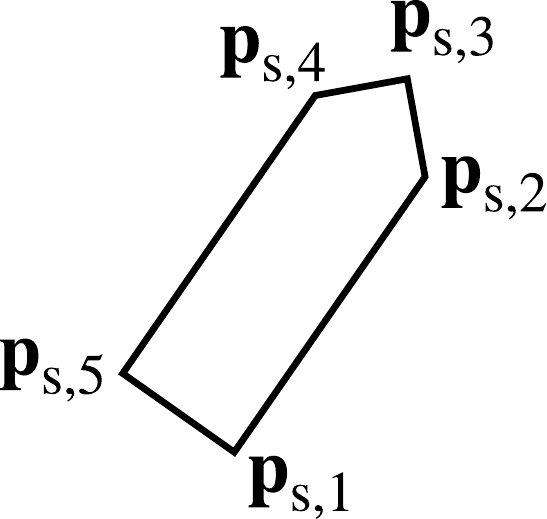}
                \vspace{8pt}
                \caption{\label{fig:fig:shipvert}}
            \end{subfigure}
            \caption{(\subref{fig:fig:boundary}) Shaded region is the obstacle-free region, formed by polygon $\Pbold\bsub$ with $\nv=11$ vertices (\subref{fig:fig:shipvert}) The points of the ship's boundary $\Pbold\ssub$ with $\ns=5$ points.}
            \label{fig:boundary}
        \end{figure}

        The docking is simulated in the Inukai experiment pond that belongs to Osaka University. The predefined obstacle-free region formed by a simple polygon $\Pbold\bsub$ is shown in Fig. \ref{fig:fig:boundary}, with number of vertices $\nv=11$. The origin of the earth-fixed coordinate system $\mathrm{o}_0-x_0y_0$ is located at $\pbold\bsubi{1}$. 
        
        The number of the points of the ship's boundary $\ns$ must be chosen carefully. Adding more points will guarantee a safer trajectory but will increase the number of constraints and the computational time. In the simulation, $\ns=5$ is considered sufficient as shown in Fig. \ref{fig:fig:shipvert}.  
        
    \subsection{On the Discretization and Interpolation}\label{subSec:3.2.-OptimSett}
        The whole trajectory is divided into $\Ns=20$ equal-spaced segments so that the number of discretization points is $\Nc=41$. Upon solving for the optimal control and states at each discretization point, it is necessary to interpolate the values between those points. Unlike in the offline OCP, here the control trajectory in a segment is not constant. 
        
        For practical consideration (mechanical limitation of the vessel's actuators), it is assumed that the control trajectory between two consecutive discretization points (half a segment) is linear. It can be computed as,
        \begin{align}\label{eq:ctrlpolyflinear}
            \onopcontrl\Hsub(t)=\onopcontrl_k + \lpare{\onopcontrl_{k+1}-\onopcontrl_k}\frac{t-t_k}{\hd}\commeq
        \end{align}
        for $t\in\ttkclosed$ where $k=1,2,\dots,\Nc-1$.
        
        The state trajectory in a segment is interpolated by \textit{natural cubic spline interpolation} method \citep{NumericalRecipeC}\@. It retains the same accuracy of Hermite cubic interpolation with $C^2$ continuity. The resulting natural cubic spline interpolant within a segment $t\in\lbra{t_k, t_{k+2}}$ can be written as,
        \begin{align}
            \begin{split}
                \onoptstate\Hsub(t)&=\onoptstate_k+\fbold_k^{*}\lpare{t-t_k}\eqskip
                &\hphantom{=}-\Bigg(3\fbold_k^{*}-4\fbold_{k+1}^{*}+\fbold_{k+2}^{*}\Bigg)\frac{\lpare{t-t_k}^2}{2\hs} \eqskip
                &\hphantom{=}+\Bigg( 2\fbold_k^{*}-4\fbold_{k+1}^{*}+2\fbold_{k+2}^{*}\Bigg)\frac{\lpare{t-t_k}^3}{3\hs^2}\;\commeq
            \end{split}
        \end{align}
        for all $k$ such that $k=2q-1$ where $q=1,2,\dots,\Ns$. Here $\fbold_{k}^{*}\equiv\fbold\lpare{\onoptstate_k,\onopcontrl_k, V, \chi}$ is the true dynamics (\ref{eq:simpledyna}) evaluated at the corresponding discretization point.

\section{Simulation Results and Evaluation}\label{Sec:4-SimRes}
    This section discusses the performance of the SO-TP. The simulations are done on a laptop with 16GB RAM and 8-core processor. 
    
    \subsection{Main Cases: Examples}
        \label{subSec:4.1.-MainCases}
        The main cases M1 to M4 are intended to give some examples of how just one almost-globally optimal and feasible solution of a slightly different problem obtained from the Off-TP can be used as a warm start for different situations in the SO-TP (adhere to the first evaluation matter), see Table \ref{tab:2-CaseInit}. Note that the Off-TP and the SO-TP differ in the OCP transcription method, the form of the objective function, and how the control trajectory is assumed within a segment.
        
        Now, to emphasize the advantages of the warm start to the SO-TP, it is relevant to show the results of the SO-TP without a warm start where a linear guess is provided, i.e., linearly changing from the initial states to the desired final states. Meanwhile, the initial guess for the control is constant: $0.5\deltamax$ and $0.5\nmax$. In addition, the initial guess of the final time $\tfpla$ is $160$ s, equal to that of the warm start.
        
        Furthermore, if $\TWSCMA$ denotes the computational time with warm start by the Off-TP and $\TNOWS$ denotes the computational time without warm start,  one can calculate the computational speedup by,
        \begin{align}
            \text{\% Comp. speedup} = \frac{\TNOWS-\TWSCMA}{\TNOWS}\times 100 \%\text{.} 
        \end{align}
        
        \subsubsection{Case M1: No Wind + Same Initial States}
        \label{subsubSec:4.1.1-CaseM1}
\begin{figure}[htbp]
            \centering
                \includegraphics[width=1\columnwidth]{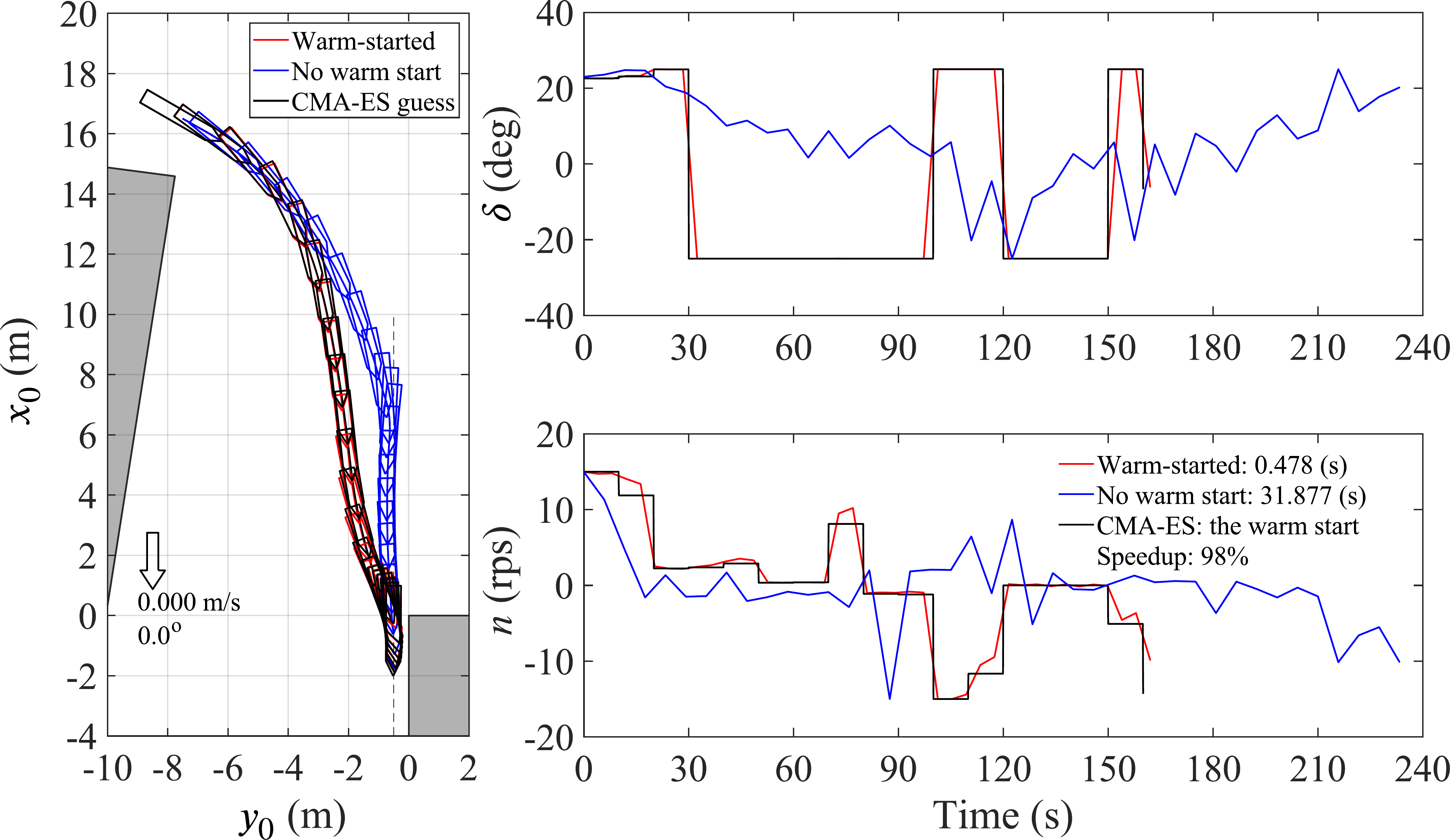}
                \caption{Case M1: summary and optimal control trajectories, no wind, with $0.478$ s calculation time.} 
                \label{fig:CaseM1ctrl}
            \end{figure}

\begin{figure}[!htbp]
            \centering
                \includegraphics[width=1\columnwidth]{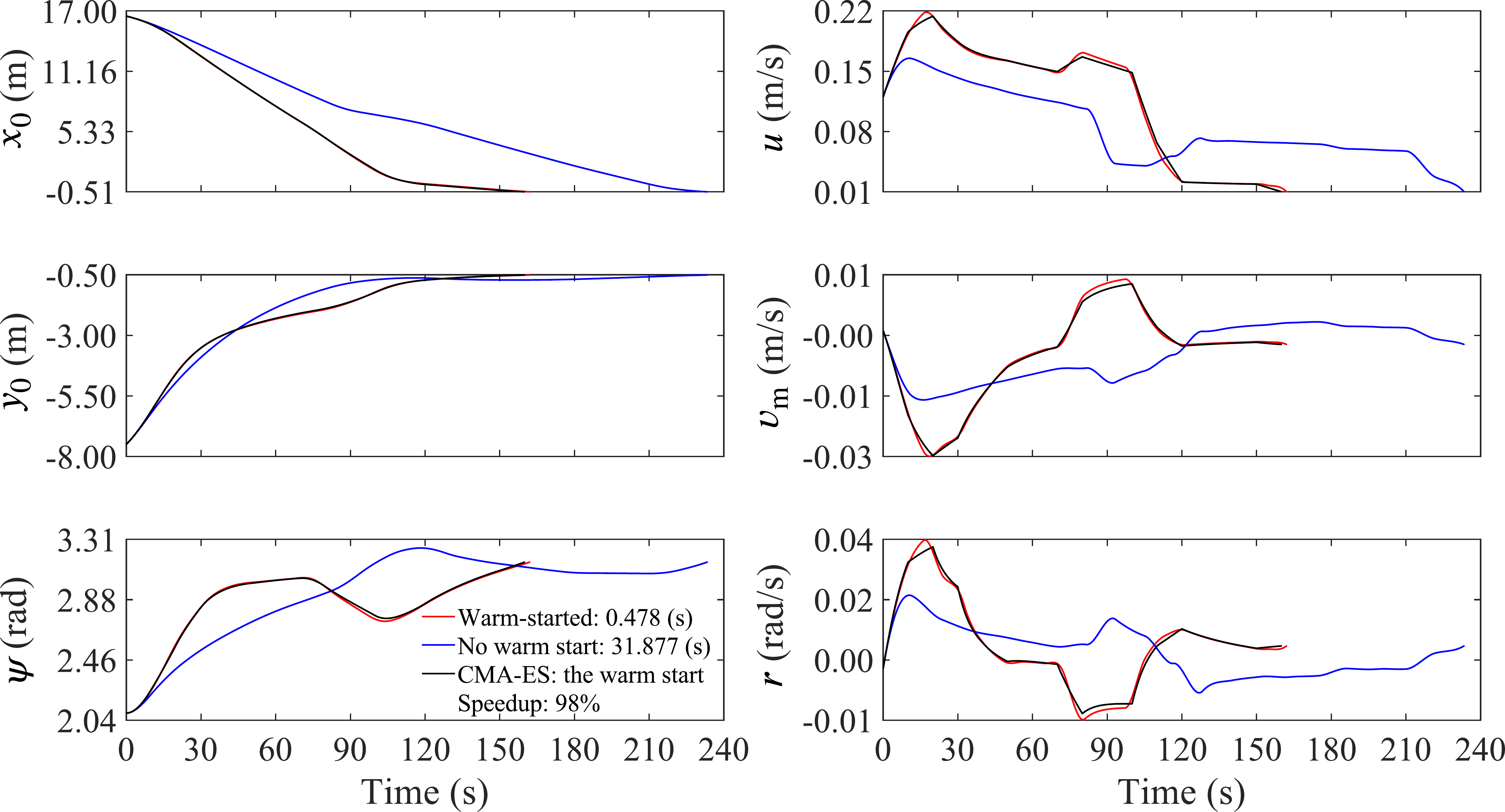}
                \caption{Case M1: optimal state trajectories, no wind, with $0.478$ s calculation time.} 
                \label{fig:CaseM1stat}
            \end{figure}

            This case is intended to confirm whether the SO-TP converges to the same warm start or not, with the same condition in which the warm start was solved, i.e., no wind and same initial states (see section \ref{Sec:3-SimulationConditions}). The comparison between the optimal control and state trajectories obtained by the SO-TP with and without a warm start for this case are shown in Fig. \ref{fig:CaseM1ctrl} and Fig. \ref{fig:CaseM1stat}. In the figures, the CMA-ES guess means the warm start from the Off-TP.
                 
            The SO-TP converged to the solution that is almost similar to the provided warm start in a relatively fast computational time. This is a clear and direct demonstration that despite the OCP in the Off-TP is formulated differently in almost all aspects, its solution belongs to the feasible set of the OCP in the SO-TP (\ref{eq:Jon}).
    
        \subsubsection{Case M2: Wind Blows Away from the Berth + Farther Initial States}
        \label{subsubSec:4.1.2-CaseM2}
\begin{figure}[htbp]
            \centering
                \includegraphics[width=\columnwidth]{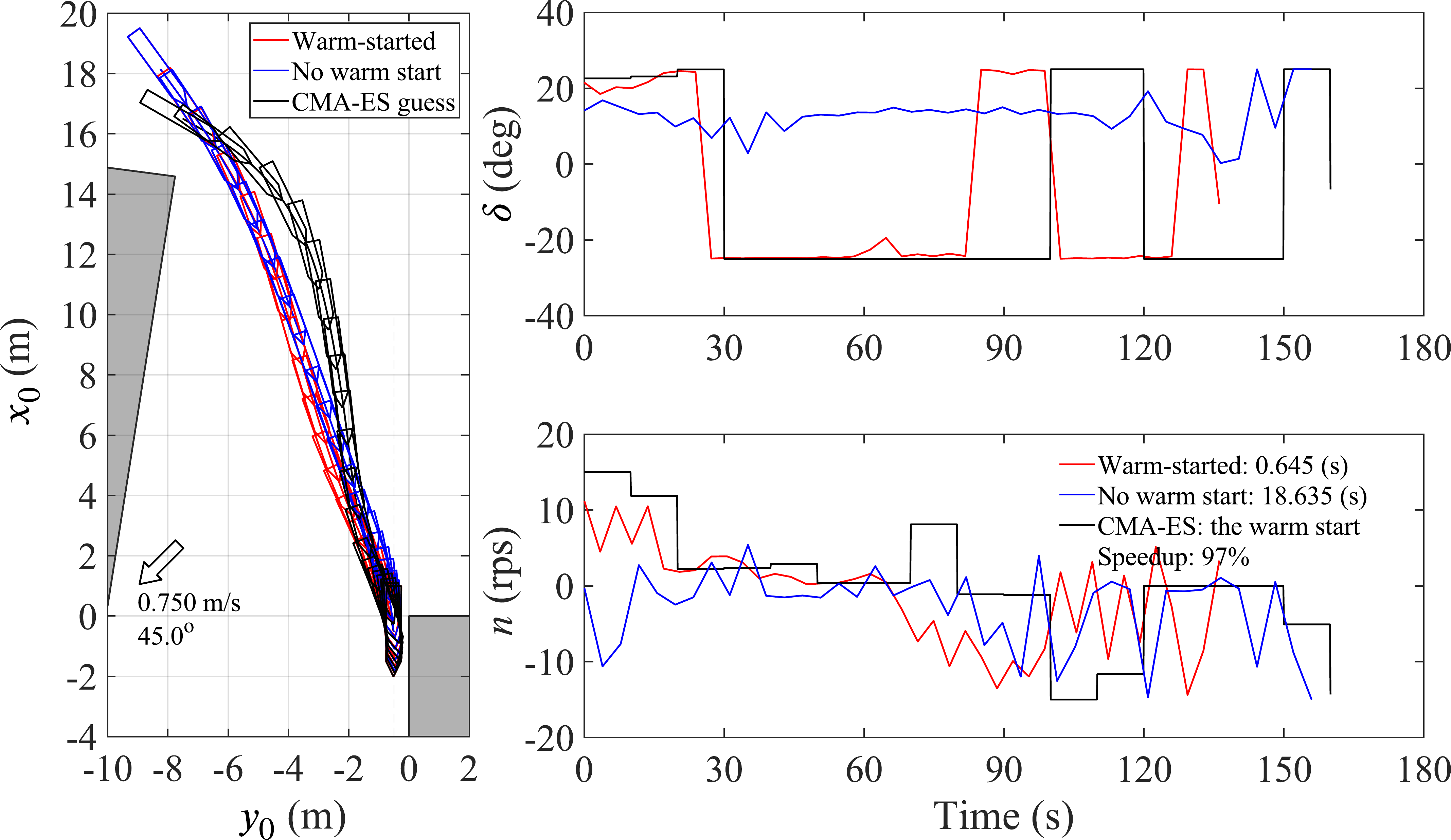}
                \caption{Case M2: summary and optimal control trajectories, $V\actutjpare=0.75$ m/s, $\chi\actutjpare=45^{\circ}$, $u\actutjpare=0.24$ m/s, $\psi\actutjpare=4\pi/5$, with $0.645$ s calculation time.} 
                \label{fig:CaseM2ctrl}
            \end{figure}

\begin{figure}[!htbp]
            \centering
                \includegraphics[width=\columnwidth]{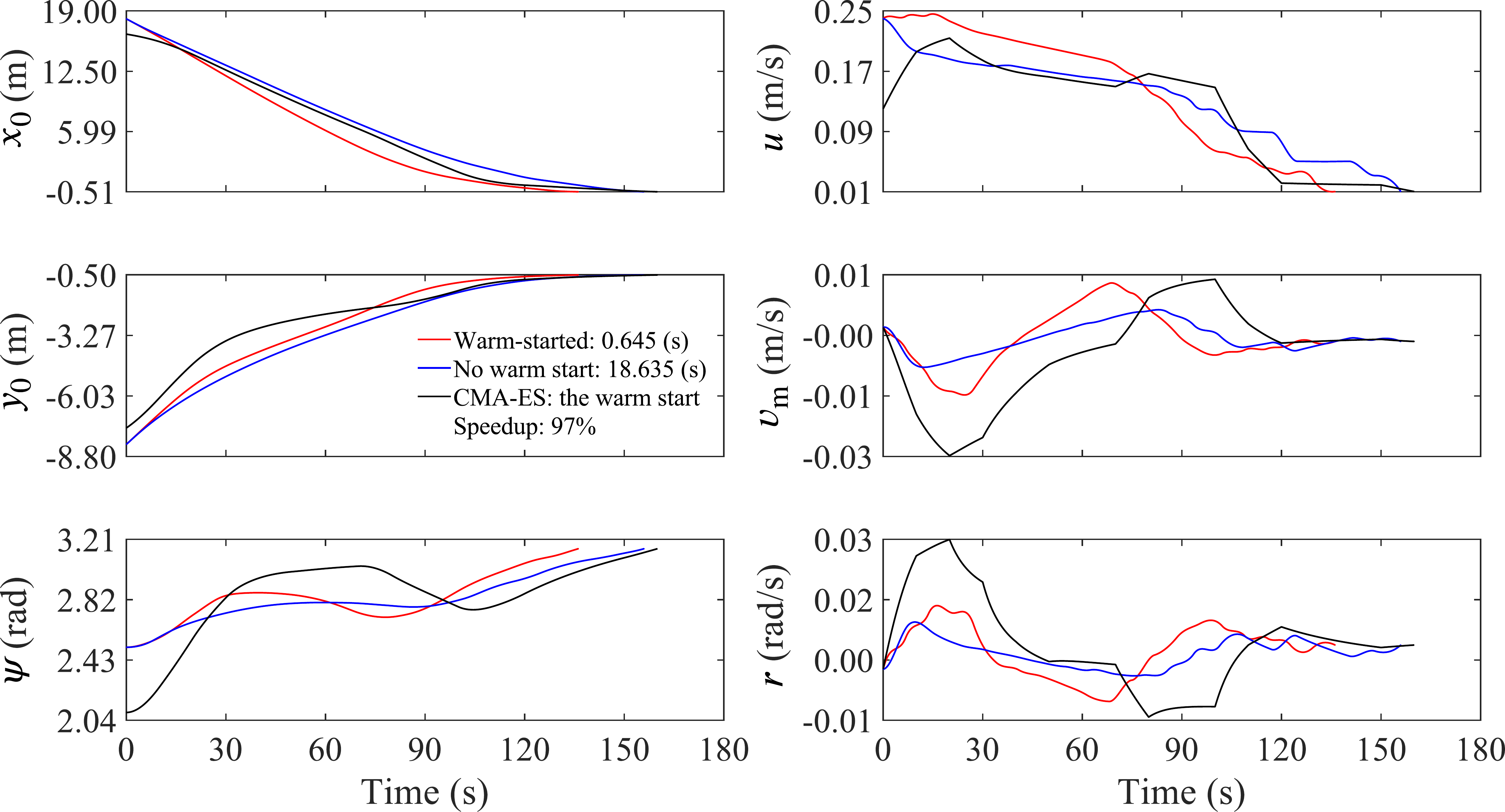}
                \caption{Case M2: optimal state trajectories, $V\actutjpare=0.75$ m/s, $\chi\actutjpare=45^{\circ}$, $u\actutjpare=0.24$ m/s, $\psi\actutjpare=4\pi/5$, with $0.645$ s calculation time.} 
                \label{fig:CaseM2stat}
            \end{figure}

            In this case, the SO-TP is executed with initial positions farther from the berth. The initial surge velocity is doubled $u(0)=2\uonint$ and the initial yaw angle is less-sharp against the berth. The wind blows at $0.75$ m/s from $45$ degree, away from the berth. The resulting optimal control and state trajectories are shown in Fig. \ref{fig:CaseM2ctrl} and Fig. \ref{fig:CaseM2stat}. 
            
            
            Because now the initial surge velocity is doubled ($0.24$ m/s; equivalent to $4.86$ knots for the full scale ship), the time $\tfpla$ that is required to reach the desired docking states $\xfin$ is shorter, even when the initial position is farther. The solution from the warm-started SO-TP gives a shorter $\tfpla$ than that without a warm start.
            
            The warm-started SO-TP also converged to a feasible and collision-free solution in a much faster computational time than that without a warm start. This M2 case demonstrates that the warm start is still applicable to a situation which initial states are different $\lpare{\text{norm }L=1.03}$ and with wind disturbance. 

        \subsubsection{Case M3: Wind Blows to the Berth + Closer Initial States}
        \label{subsubSec:4.1.3-CaseM3}    
\begin{figure}[htbp]
            \centering
                \includegraphics[width=\columnwidth]{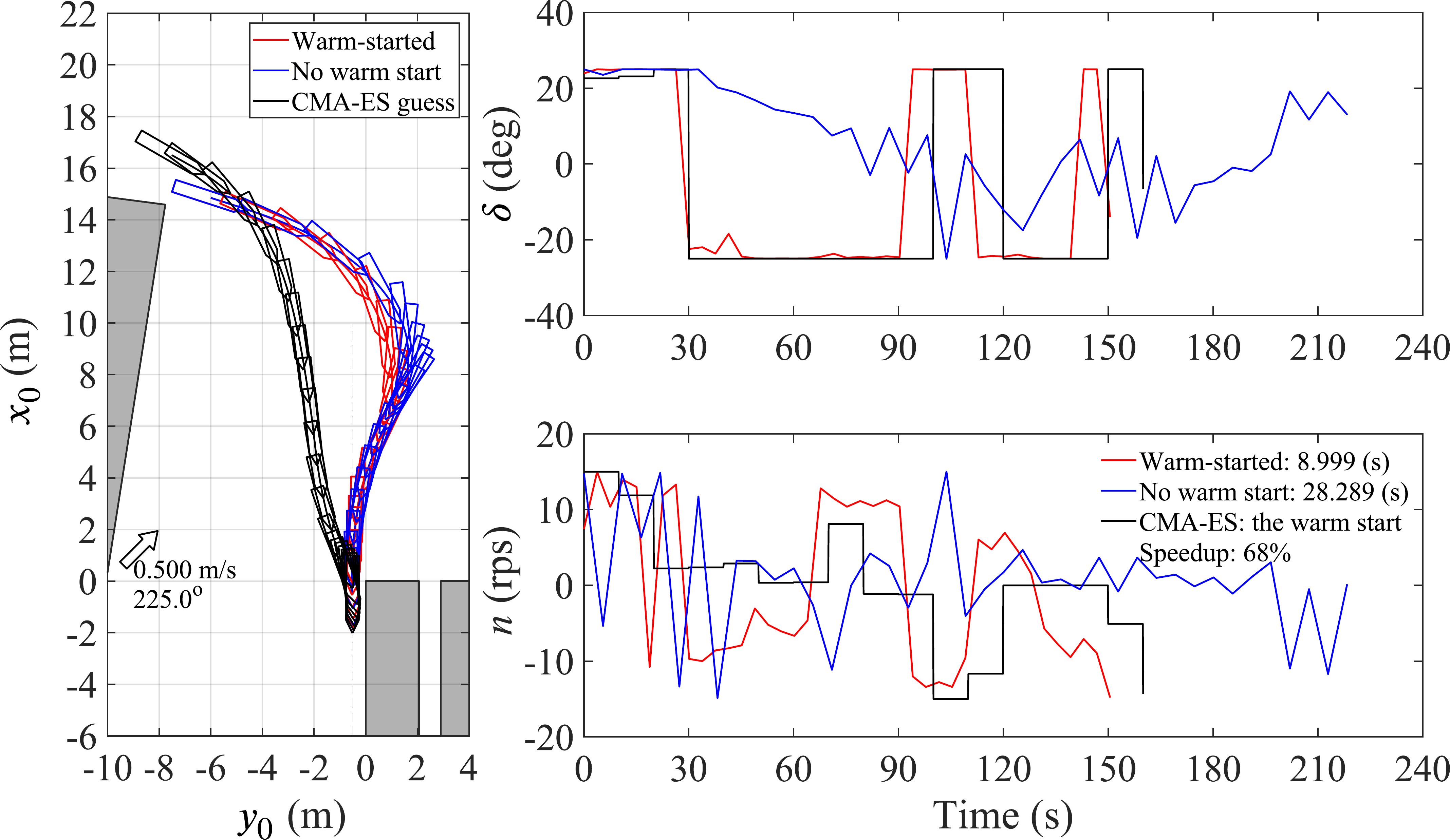}
                \caption{Case M3: summary and optimal control trajectories, $V\actutjpare=0.50$ m/s, $\chi\actutjpare=225^{\circ}$, $u\actutjpare=0.24$ m/s, $\psi\actutjpare=3\pi/5$, with $8.999$ s calculation time.} 
                \label{fig:CaseM3ctrl}
            \end{figure}

\begin{figure}[!htbp]
            \centering
                \includegraphics[width=\columnwidth]{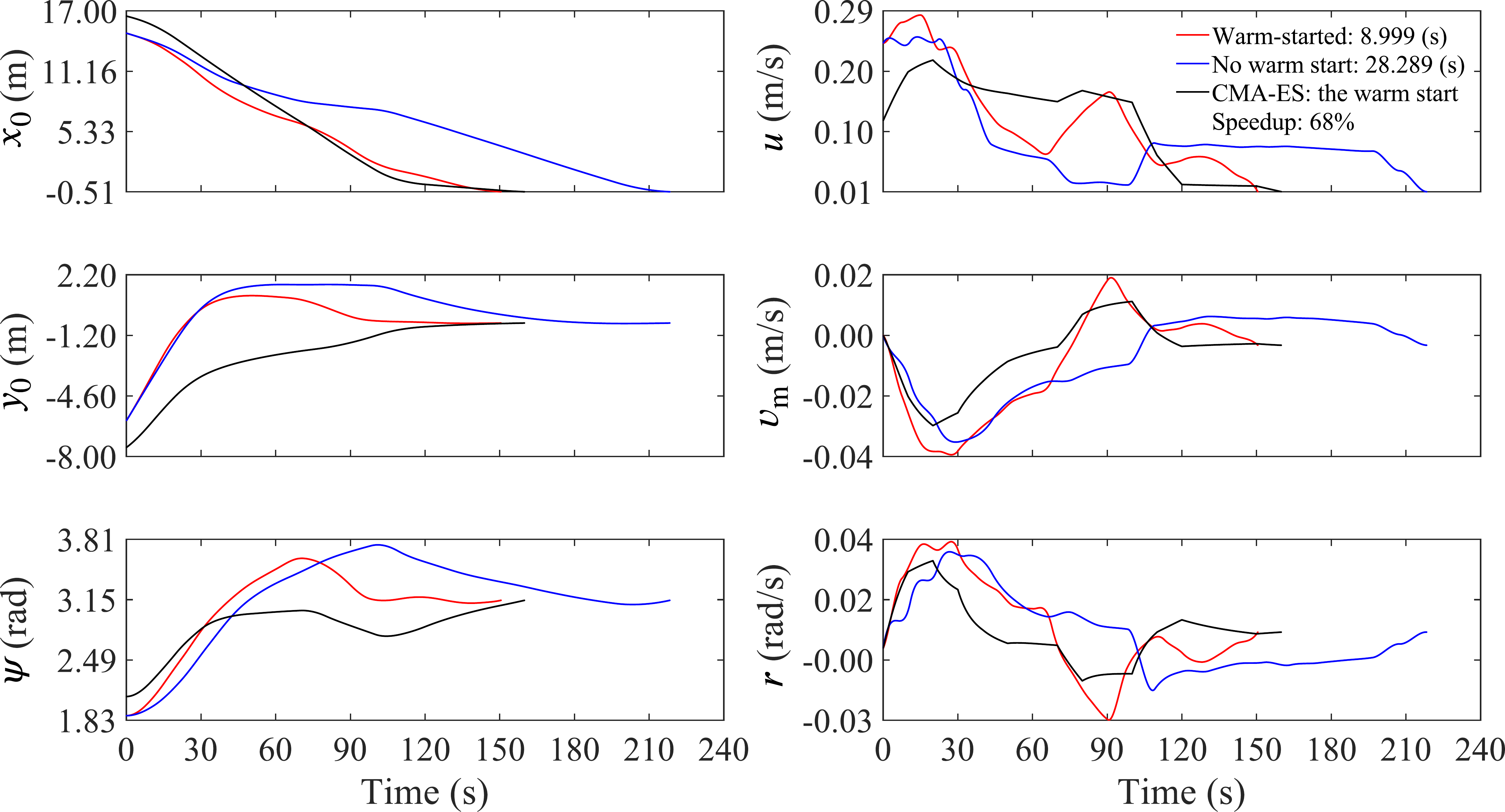}
                \caption{Case M3: optimal state trajectories, $V\actutjpare=0.50$ m/s, $\chi\actutjpare=225^{\circ}$, $u\actutjpare=0.24$ m/s, $\psi\actutjpare=3\pi/5$, with $8.999$ s calculation time.} 
                \label{fig:CaseM3stat}
            \end{figure}

            In contrast to the previous case M2, the initial position of the ship in case M3 is closer to the berth and the initial yaw angle is sharper against the berth. Such initial condition limits the free area for the ship to maneuver. In this case, the wind speed is $0.5$ m/s and it blows from at $225$ degree quarterly to the berth. The resulting optimal control and state trajectories for this case are visualized in Fig. \ref{fig:CaseM3ctrl} and Fig. \ref{fig:CaseM3stat}. 
            
            
            It can be seen that now the problem becomes quite difficult for the ship to reach the desired docking states $\xfin$ form the given initial states because the ship is expected to do a sharp turn in such a relatively narrow area. This is implied by the longer computational time that is required for the SO-TP to converge. Nevertheless, the warm start gives a considerable $68$\% computational speedup.
            
            Similar to the previous case, $\tfpla$ is shorter than that of the warm start, this is because the initial surge velocity is doubled; small thrust from the propeller can give more momentum for the ship to do a sharp turn. From Fig. \ref{fig:CaseM3ctrl}, one can also observe that the SO-TP without a warm start results in $\tfpla$ that is much longer than the warm-started SO-TP.
            
            Keen readers may find it interesting to see the same tendency that the resulting optimal $\delta$ trajectory is similar to the warm start. This shows that the thrust from the propeller contributes significantly to the overall maneuverability of the ship. An intuitive explanation to this tendency is that the warm start for $\delta$ gives the objective function value that is in the neighborhood of the curvature of the interpolation (\ref{eq:intpconston}) and quadrature equality constraints (\ref{eq:quadconston}), while that for $n$ is far from the curvature of the equality constraints. 
            
        \subsubsection{Case M4: Wind Blows Away from the Berth + Large Deviation}
        \label{subsubSec:4.1.4-CaseM4}
\begin{figure}[htbp]
            \centering
                \includegraphics[width=\columnwidth]{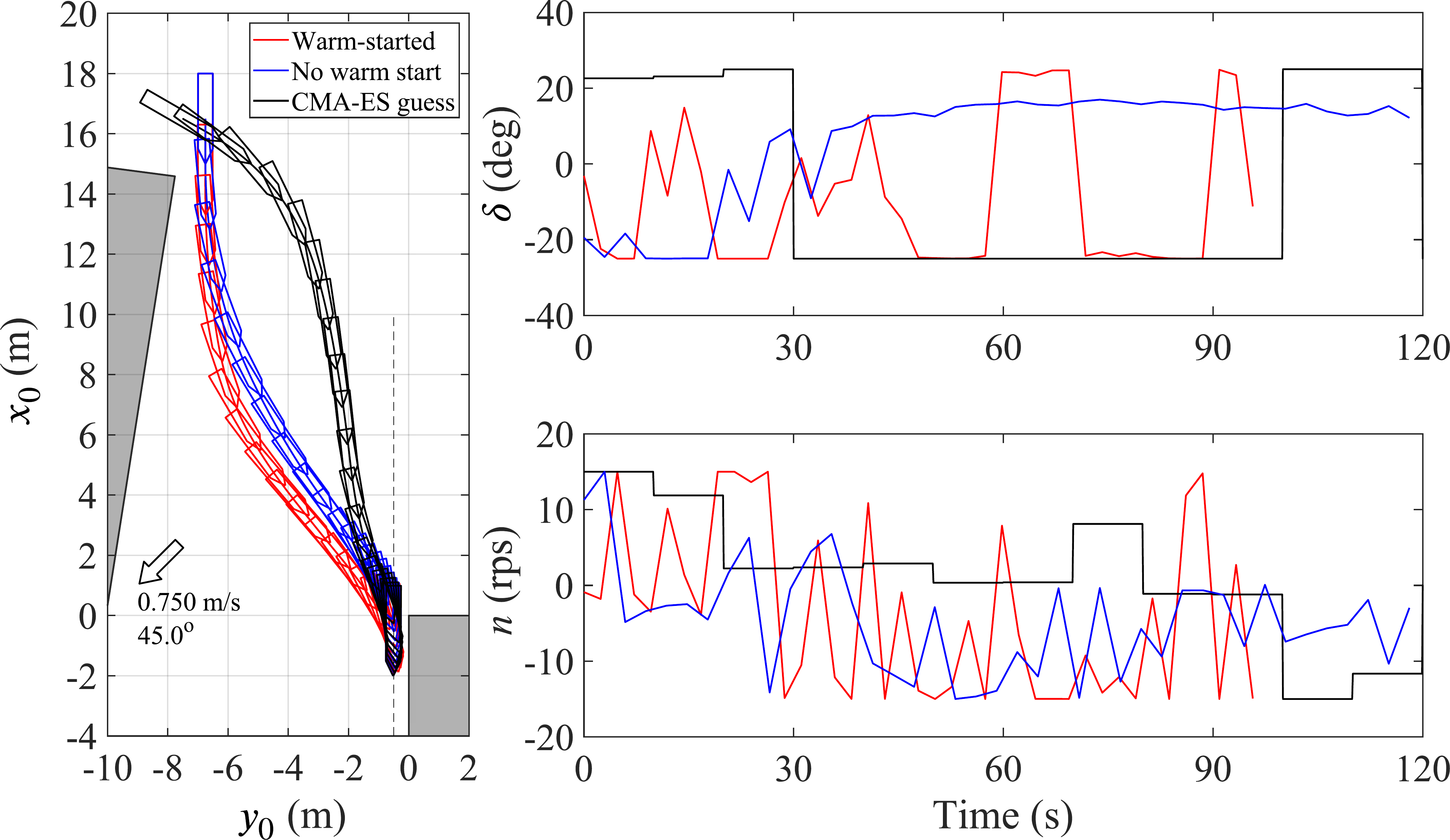}
                \caption{Case M4: summary and optimal control trajectories, $V\actutjpare=0.75$ m/s, $\chi\actutjpare=45^{\circ}$, $u\actutjpare=0.36$ m/s, $\psi\actutjpare=\pi$, with $11.664$ s calculation time.} 
                \label{fig:CaseM4ctrl}
            \end{figure}

\begin{figure}[tbp]
            \centering
                \includegraphics[width=\columnwidth]{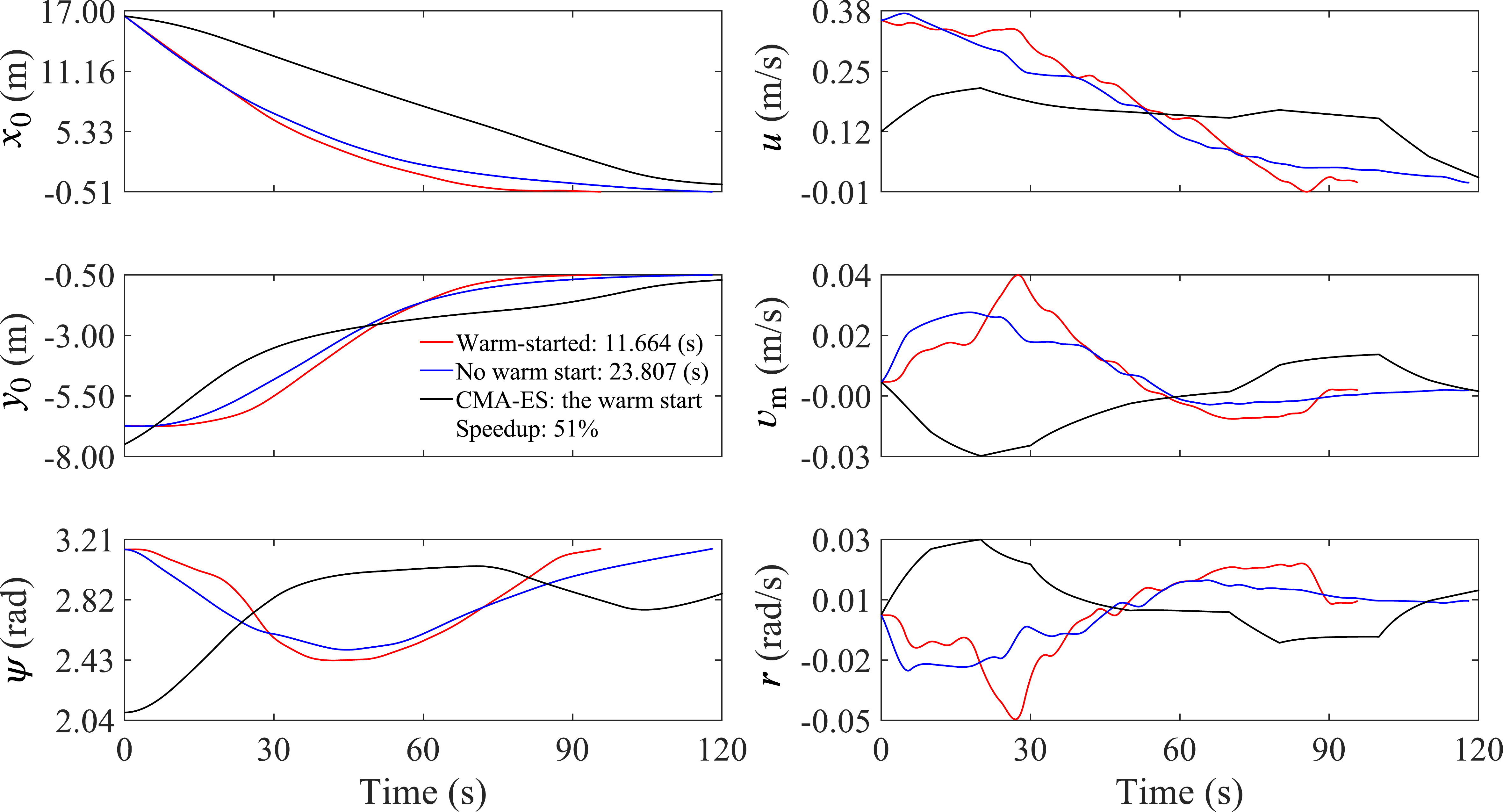}
                \caption{Case M4: optimal state trajectories, $V\actutjpare=0.75$ m/s, $\chi\actutjpare=45^{\circ}$, $u\actutjpare=0.36$ m/s, $\psi\actutjpare=\pi$, with $11.664$ s calculation time.} 
                \label{fig:CaseM4stat}
            \end{figure}

            Suppose now the ship arrives at an entry location with initial yaw angle that is quite different from that of the warm start: parallel to the berth. The ship also arrives at a surge velocity $0.36$ m/s; three times of that of the warm start: equivalent to $7.28$ knots for the full scale ship. In this case, the wind blows at $0.75$ m/s away from the berth. Such condition is very different from the warm start with norm $L=2.06$. The resulting optimal control and state trajectories are shown in Fig. \ref{fig:CaseM4ctrl} and Fig. \ref{fig:CaseM4stat}.     
            
            Despite the large difference of the initial states from the warm start, the warm-started SO-TP still gives a faster computational time than that without a warm start: $51$\% speedup. This means that the warm start is still applicable to speed up the computation, even for a situation that is very much different from that when obtaining it.
            
            To conclude this subsection about the first evaluation matter, the readers now must have a good grasp of the applicability of one almost-globally optimal solution from the Off-TP to warm-start the SO-TP for various different scenarios. The resulting trajectories are guaranteed to be collision-free and satisfy all the given constraints. 
   
    \subsection{Additional Trial Cases}
        \label{subSec:4.2-AdditionalCases}
        As to further show the effect of the warm start in the computational time and the performance of the SO-TP (adhere to the second evaluation matter), 10 additional cases (A1 to A10) are also considered (see Table \ref{tab:2-CaseInit}). The same offline solution from the Off-TP is used to warm-start the SO-TP. Note that A9 and A10 are the cases when the ship starts the docking operation from the East (positive $y_0$ axis), while the rest of the cases (M1 to A8) are from the West. For the sake of demonstration, these additional cases are chosen arbitrarily, with norm of the initial states $L$ varies between 0 to 2.3.
        
        The effect of the warm start is evaluated based on the the computational time and the resulting $\tfpla$. The summary of these additional trial cases along with the main cases is presented in Table \ref{tab:3-CompareWarmandNoWarmStart}. It can be seen that the warm start by the Off-TP gives a significant reduction in the computational time. As a comparison, the average computational time of the warm-started SO-TP is $9.211$ s while that without a warm start is $34.292$ s. The warm start by just one almost-globally optimal solution from the Off-TP gives an average of $71\%$ computational time speedup. In addition, the warm-started SO-TP gives a shorter $\tfpla$ for almost all of the cases, which is desired.
        
        \begin{table*}\centering
        
        \caption{Comparison of the SO-TP: with and without a warm start (16 GB RAM, 8-core processor laptop). See Fig. \ref{fig:App1} to Fig. \ref{fig:App3} in the appendix.}
        \label{tab:3-CompareWarmandNoWarmStart}       % Give a unique label
        % For LaTeX tables use
            \begin{tabular}{cccc|cc|cc|c}
                \hline\noalign{\smallskip}
                
                \multirow{3}{*}{Case} & Norm $L=$ & $V\actutjpare$ & $\chi\actutjpare$ & \multicolumn{2}{c|}{Warm start by Off-TP} & \multicolumn{2}{c|}{No warm start (linear)} & \multirow{2}*{$\%$ Comp. speedup} \\
                
                \noalign{\smallskip}\cline{5-8}\noalign{\smallskip}
                & \multirow{2}*{$\sqrt{\sum_i^{6}\lpare{d_i - 1}^2}$} & \multirow{2}*{(m/s)} & \multirow{2}*{(deg)} & \multirow{2}{*}{$\tfpla$ (s)}  & Comp. time& \multirow{2}{*}{$\tfpla$ (s)} & Comp. time& \multirow{2}{*}{ $\frac{\TNOWS-\TWSCMA}{\TNOWS}\times 100\%$}\\
                
                & & & & & $\TWSCMA$ (s) &  & $\TNOWS$ (s) & \\
                
                \noalign{\smallskip}\hline\noalign{\smallskip}
                M1 & $0.00$ & \multicolumn{2}{c|}{No wind} & $162.0$ & $0.478$ & $233.3$ & $31.877$ & $98\%$\\
                
                M2 & $1.03$ & $0.75$ & $45$ & $136.2$ & $0.645$ & $156.0$ & $18.635$ & $97\%$\\
                
                M3 & $1.03$ & $0.50$ & $225$ & $150.5$ & $8.999$ & $218.4$ & $28.289$ & $68\%$\\
                
                M4 & $2.06$ & $0.75$ & $45$ & $95.7$ & $11.664$ & $118.1$ & $23.807$ & $51\%$\\
                
                A1 & $0.32$ & $0.75$ & $45$ & $151.9$ & $3.425$ & $177.1$ & $18.361$ & $81\%$\\
                
                A2 & $0.53$ & $0.75$ & $180$ & $163.1$ & $5.881$ & $231.8$ & $17.711$ & $67\%$\\
                
                A3 & $0.79$ & $0.75$ & $125$ & $146.5$ & $20.718$ & $166.1$ & $46.836$ & $56\%$\\
                
                A4 & $0.92$ & $0.50$ & $135$ & $167.7$ & $3.426$ & $289.4$ & $47.019$ & $93\%$\\
                
                A5 & $0.58$ & $0.50$ & $90$ & $175.5$ & $6.084$ & $217.5$ & $31.397$ & $81\%$\\
                
                A6 & $0.41$ & $0.50$ & $315$ & $159.6$ & $3.442$ & $256.1$ & $54.906$ & $94\%$\\    
                
                A7 & $1.25$ & $0.50$ & $250$ & $207.0$ & $14.313$ & $205.9$ & $19.903$ & $28\%$\\ 
                
                A8 & $1.51$ & $0.50$ & $90$ & $137.4$ & $9.268$ & $159.7$ & $68.519$ & $86\%$\\
                
                A9 & $2.24$ & $0.75$ & $0$ & $180.5$ & $17.487$ & $214.5$ & $30.697$ & $43\%$\\
                
                A10 & $2.31$ & $0.50$ & $45$ & $203.7$ & $23.128$ & $192.7$ & $42.133$ & $45\%$\\
                \noalign{\smallskip}\hline\noalign{\smallskip}
                
                & & & & AVG & $9.211$ & AVG & $34.292$ & $71\%$\\
                \noalign{\smallskip}\cline{5-9}
            \end{tabular}
        \end{table*}    
        
       \begin{table*}\centering
        \caption{Comparison of the SO-TP: warm-started by the Off-TP and warm-started by the SO-TP for case M1 with no warm start (linear guess).}
        \label{tab:4-CompareWarmStart}      
            \begin{tabular}{c|cccc|cccc|c}
                \hline\noalign{\smallskip}
                
                \multirow{3}{*}{Case} & \multicolumn{4}{c|}{Warm start by Off-TP} & \multicolumn{4}{c|}{Warm start by SO-TP (M1, no warm start)} & \multirow{2}*{Preferred} \\
                
                \noalign{\smallskip}\cline{2-9}\noalign{\smallskip}
                & \multirow{2}{*}{$\tfpla$ (s)} & Feasi- & Constraint & Comp. time & \multirow{2}{*}{$\tfpla$ (s)} & Feasi- & Constraint & Comp. & warm start \\
                
                & & ble? & violation & $\TWSCMA$ (s) & &  ble? & violation & time (s) & (based on comp. time)\\
                
                \noalign{\smallskip}\hline\noalign{\smallskip}
                
                M2 & $136.2$ & Yes & $2.5\mathrm{E}-11$ & $0.645$ & $-$ & No & $2.1\mathrm{E}-3$ & $77.111$ & Off-TP\\
                
                M3 & $150.5$ & Yes & $2.5\mathrm{E}-11$ & $8.999$ & $230.7$ & Yes & $1.3\mathrm{E}-11$ & $6.094$ & SO-TP\\
                
                M4 & $95.7$ & Yes & $1.7\mathrm{E}-11$ & $11.664$ & $-$ & No & $8.3\mathrm{E}-3$ & $110.403$ & Off-TP\\
                
                A1 & $151.9$ & Yes & $4.5\mathrm{E}-12$ & $3.425$ & $172.1$ & Yes & $6.7\mathrm{E}-13$ & $17.657$ & Off-TP\\
                
                A2 & $163.1$ & Yes & $2.2\mathrm{E}-11$ & $5.881$ & $245.3$ & Yes & $5.4\mathrm{E}-13$ & $12.682$ & Off-TP\\
                
                A3 & $146.5$ & Yes & $1.4\mathrm{E}-11$ & $20.718$ & $-$ & No & $2.2\mathrm{E}-3$ & $161.069$ & Off-TP\\
                
                A4 & $167.7$ & Yes & $2.0\mathrm{E}-11$ & $3.426$ & $-$ & No & $3.1\mathrm{E}-3$ & $37.161$ & Off-TP\\
                
                A5 & $175.5$ & Yes & $2.4\mathrm{E}-12$ & $6.084$ & $-$ & No & $1.3\mathrm{E}-3$ & $66.414$ & Off-TP\\
                
                A6 & $159.6$ & Yes & $5.2\mathrm{E}-12$ & $3.442$ & $-$ & No & $3.2\mathrm{E}-3$ & $47.519$ & Off-TP\\    
                
                A7 & $207.0$ & Yes & $9.7\mathrm{E}-11$ & $14.313$ &  $248.4$ & Yes & $1.2\mathrm{E}-11$ & $27.535$ & Off-TP\\ 
                
                A8 & $137.4$ & Yes & $5.9\mathrm{E}-12$ & $9.268$ & $152.1$ & Yes & $1.1\mathrm{E}-11$ & $297.179$ & Off-TP\\
                
                A9 & $180.5$  & Yes & $9.7\mathrm{E}-14$ & $17.487$ & $210.1$ & Yes & $3.3\mathrm{E}-11$ & $44.545$ & Off-TP\\
                
                A10 & $203.7$ & Yes & $4.0\mathrm{E}-12$ & $23.128$ & $325.5$ & Yes & $6.4\mathrm{E}-11$ & $29.416$ & Off-TP\\
                    
                \noalign{\smallskip}\hline
            \end{tabular}
        \end{table*}           
        
    \subsection{The Necessity for the Off-TP}
        \label{subSec:4.3-NecessityOff-TP}
         In adherence to the last evaluation matter, the authors tried to use the locally optimal solution for case M1 obtained by the SO-TP without a warm start (use a linear initial guess, see Fig. \ref{fig:CaseM1ctrl} and Fig. \ref{fig:CaseM1stat}) to warm-start the SO-TP for the rest of the cases (M2 to A10). 
         
         Following the previous subsection, the computational time and the resulting $\tfpla$ are used as comparison between the two warm start methods (by the Off-TP and by the SO-TP). In addition, the feasibility status is also reported. This feasibility is determined by how much the optimization variables violate the equality constraints described in (\ref{eq:initstaton}) to (\ref{eq:sumofangconst}). Infeasibility is reported if one or more variables result in a constraint violation larger than the default tolerance value of $1\mathrm{E}-6$. The summary is presented in Table \ref{tab:4-CompareWarmStart}.
        
        From Table \ref{tab:4-CompareWarmStart}, almost half of the cases have one or more optimization variables that violate the constraints tolerance: infeasible. It is apparent that the solution from the SO-TP without a warm start should not be used as an initial guess for different problems as it can not dictate the search to a feasible solution. This situation implies the nonlinear nature of the NLP and that the SQP requires a good "starting point" for the search to converge to a feasible solution.
        
        Therefore, to maximize the potential of the SO-TP, a good initial guess is important to warm-start the optimization. Such initial guess can readily be obtained by the Off-TP which applicability to many situations have been demonstrated in subsection \ref{subSec:4.1.-MainCases} and summarized in Table \ref{tab:3-CompareWarmandNoWarmStart}. To conclude this subsection, the Off-TP and SO-TP should be implemented together so that both can exploit their respective advantages to compromise each other's disadvantages.
            
\section{Concluding Remarks}\label{Sec:5-ConcludeRemarks}        
    This article demonstrates a warm-started semionline trajectory planner (SO-TP) that is applicable to underactuated vessels, able to guarantee that the feasible trajectory is safe, and able to consider the wind dynamics up to some extent. It has been shown that significant reduction in the computational time of the SO-TP can be achieved with the solution from the offline trajectory planner (Off-TP) as a warm start. The versatility of just one solution from the Off-TP to warm-start the SO-TP for many different scenarios have also been demonstrated. 
    
    However, there is no guarantee that the warm start or any initial guess will result in a feasible solution or computational time speedup. The underactuation and the restricted harbor area greatly limit the existence of the feasible solutions, which is reasonable. This suggests that the presence of a feedback controller is necessary, so that practically when the trajectory planner fails, the previously obtained trajectory can still be used as the set point for the feedback controller.
    
    The semionline (not in real time) nature of the computational time is a direct consequence from the complexity of the problem and the underactuation of the system: a harder NLP and more mathematical function evaluations. Nevertheless, frequently replanning the reference trajectory may lead to the instability of the feedback controller. Moreover, the average computational time from the 14 trial cases shown in this article is less than 30 seconds, which is relatively fast for a full scale ship. For this reason, this article can be regarded as a contribution to realize an online trajectory planner for an underactuated vessel in the future.  
    
    As an idea for the next work, one can construct a database of offline solution by the Off-TP from several initial states and wind conditions. These offline solutions can be organized into a grid system from which the SO-TP will choose one best offline solution to be used as its warm start. This is why the authors introduced the norm $L$ in the discussion: to measure the deviation of the initial states from that of the offline solution, i.e., chose one offline solution that gives the smallest $L$ as the warm start. By doing this, given by the demonstrated versatility of one solution from the Off-TP to speed up the SO-TP for many different situations, it is more likely that the average computational time can be further shortened. This will be implemented in the near future in an actual experiment with the model scale.
    
    Even though not related to autonomous ship, as an example of the implementation of the warm-started SO-TP, a shipmaster can use the resulting trajectory as a guide when maneuvering the ship to dock. Moreover, the feasibility status from the SO-TP can be used as a qualitative measure as to whether the shipmaster should proceed to dock or not. In this sense, when the SO-TP can not give a feasible solution, within the accuracy of the mathematical model of the ship, it is safe to say that the ship can not reach the desired docking states from the given initial states under a certain condition by itself.
    
% Uncomment and use as the case may be
%\begin{theorem} 
%\end{theorem}

% Uncomment and use as the case may be
%\begin{lemma} 
%\end{lemma}

%% The Appendices part is started with the command \appendix;
%% appendix sections are then done as normal sections
\appendix

\section*{Declaration of Competing Interests}
The authors declare that they have no known competing financial interests or 
personal relationships that could have appeared to influence the work reported in 
this paper.

\section*{Acknowledgment}\label{Acknowledgment}  
This study was supported by a Grant-in-Aid for Scientific Research from the Japan Society for Promotion of Science (JSPS KAKENHI Grant \#19K04858).

% To print the credit authorship contribution details
\printcredits

\section{Appendix}\label{Sec:6-Appendix}   
    \setcounter{figure}{0} 
    \setcounter{table}{0} 
    \renewcommand{\thetable}{\thesection.\arabic{table}}
    \renewcommand{\thefigure}{\thesection.\arabic{figure}}
        \begin{table}[pos=h]
        \caption{Principal particulars of Esso Osaka 1/108 model scale.}\label{tab:A1-EssoOsakaPP}
            \begin{tabular*}{\tblwidth}{@{}LC@{}}
                \toprule
                 Parameters & Measurement\\ 
                \midrule
                Length between perpendiculars $L_{\mathrm{pp}}$ (m) & $3.000$  \\
                Breadth $B$ (m) & $0.489$  \\
                Draught $d$ (m) & $0.201$ \\
                Longitudinal center of gravity $\xG$ (m) & $0.095$ \\
                Mass $m$ (kg) & $245.091$ \\
                Block coefficient $C_{\mathrm{b}}$ & $0.831$ \\
                \bottomrule
            \end{tabular*}
        \end{table} 
        
        \begin{figure}[!h]
            \centering
                \includegraphics[width=0.6\columnwidth]{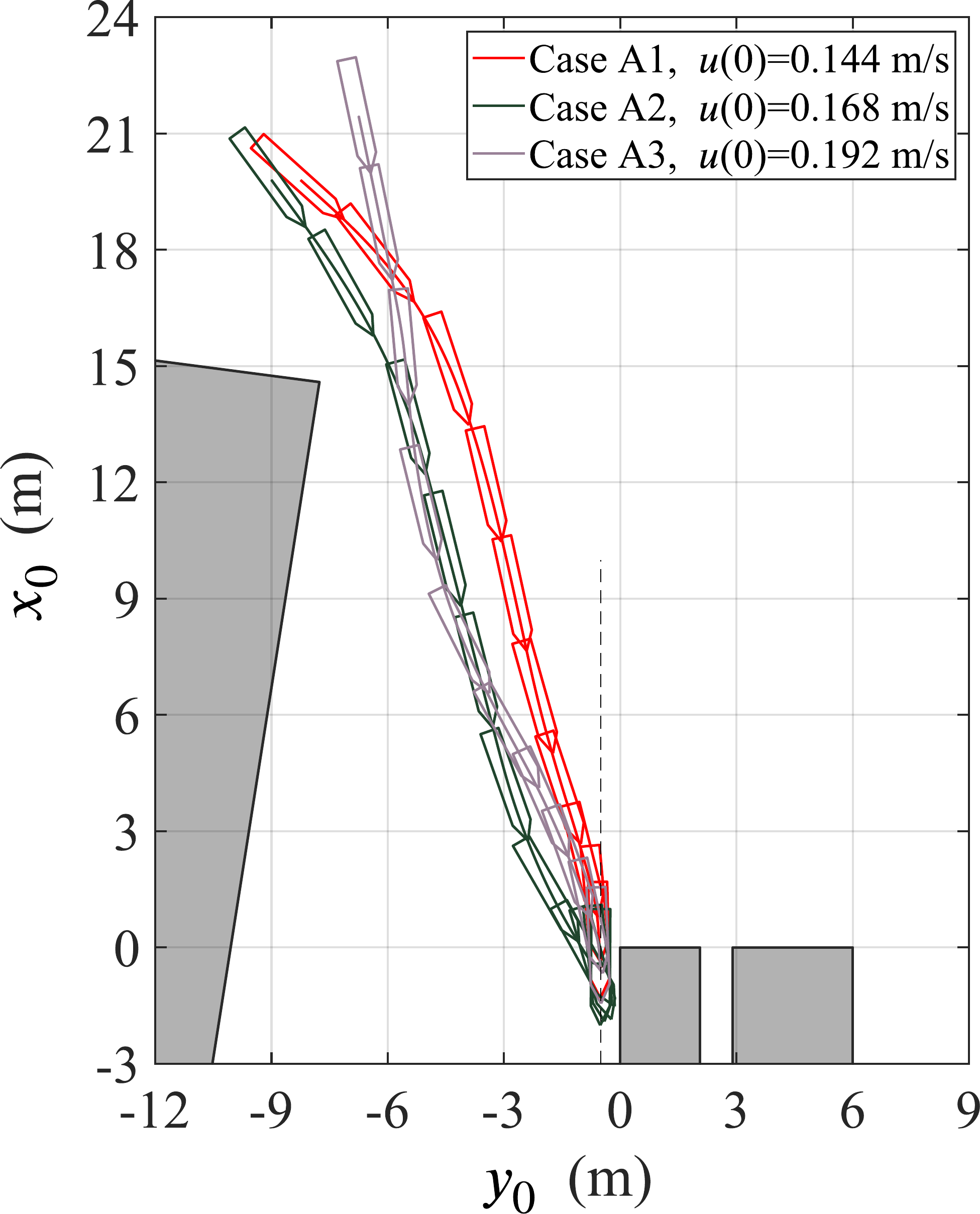}
                \caption{Result from the warm-started SO-TP: case A1 to case A3.} 
                \label{fig:App1}
        \end{figure}    
        
        \begin{figure}[!h]
            \centering
                \includegraphics[width=0.6\columnwidth]{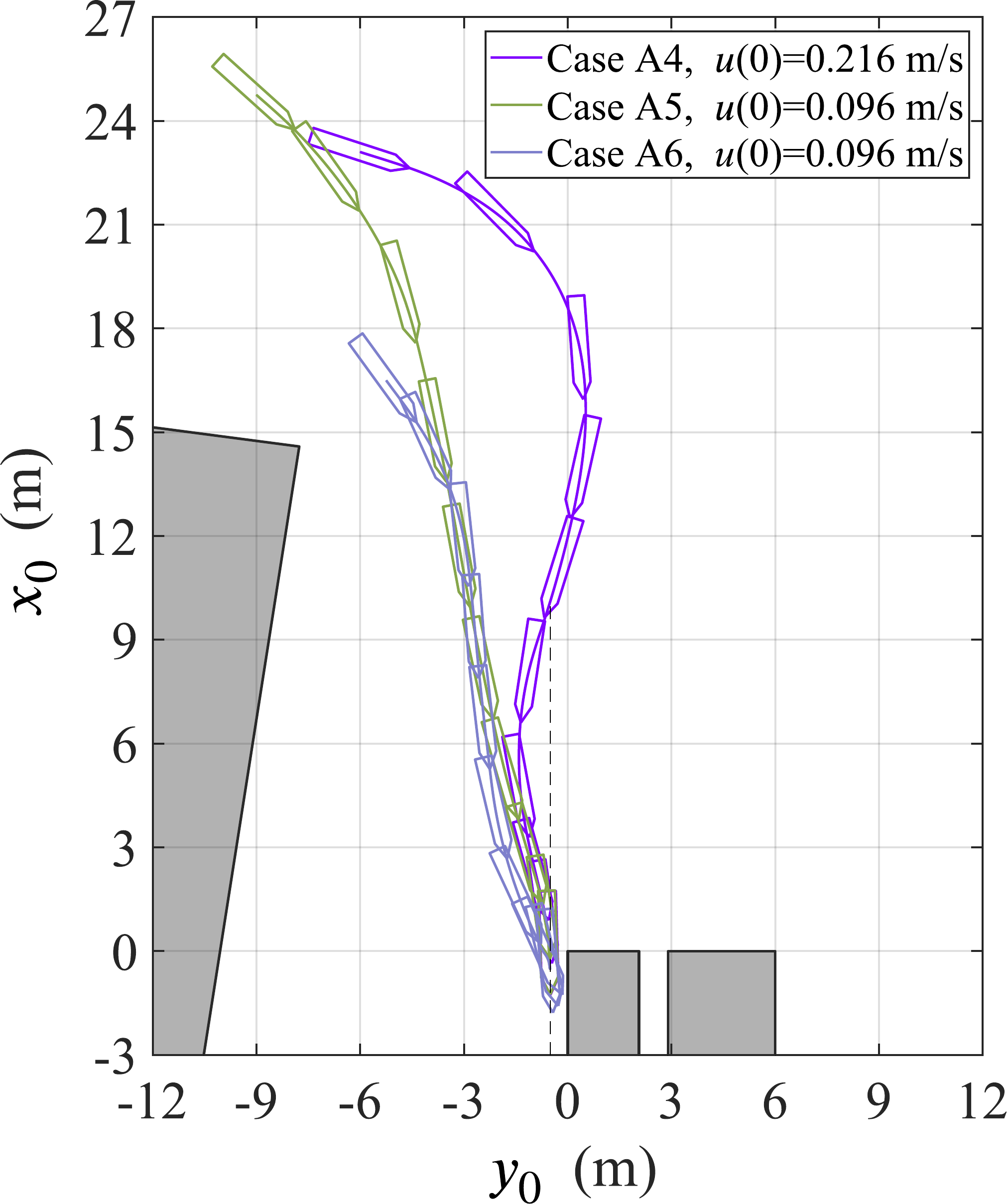}
                \caption{Result from the warm-started SO-TP: case A4 to case A6.} 
                \label{fig:App2}
        \end{figure}  
        
        \begin{figure}[!h]
            \centering
                \includegraphics[width=0.6\columnwidth]{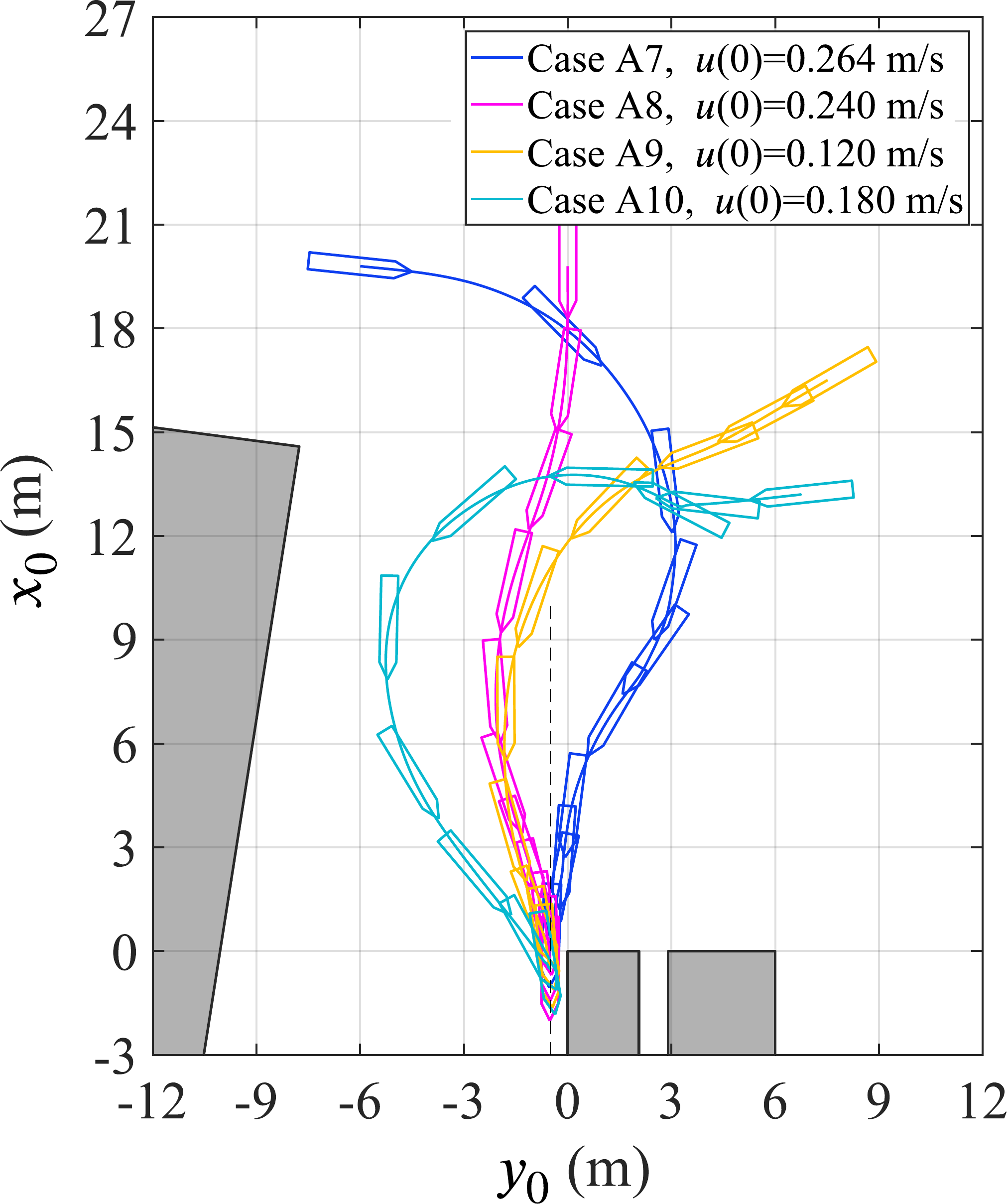}
                \caption{Result from the warm-started SO-TP: case A7 to case A10.} 
                \label{fig:App3}
        \end{figure}          

%% Loading bibliography style file
%\bibliographystyle{model1-num-names}
%\bibliographystyle{cas-model2-names}

% Loading bibliography database
%\bibliography{bibsource}

% Biography
%\bio{}
% Here goes the biography details.
%\endbio

%\bio{pic1}
% Here goes the biography details.
%\endbio

\end{document}